\documentclass[referee]{raa} 
\usepackage{lscape}           
\usepackage{natbib}
\usepackage[a4paper, left=25mm, right=25mm, top=20mm, bottom=20mm]{geometry}
\bibstyle{model2-names.bst}
\usepackage{graphicx,times}             
\usepackage{amssymb,amsmath}
\bibpunct{(}{)}{;}{a}{}{,}

\usepackage[a4paper=true,dvipdfm=true,pagebackref=true]{hyperref}
\hypersetup{colorlinks = true, linkcolor = green, anchorcolor = red, citecolor = blue, filecolor = red, pagecolor = red, urlcolor = red}

\begin{document}

   \title{Analysis of ground level enhancement events of 29 September 1989; 15 April 2001 and 20 January 2005
}

 \volnopage{Vol.0 (20xx) No.0, 000--000}      
   \setcounter{page}{1}          

   \author{R.E. Ugwoke
      \inst{1,2*}
      \and A. A. Ubachukwu
      \inst{2}
      \and J. O. Urama
      \inst{2}
      \and O. Okike
      \inst{3}
      \and J. A. Alhassan
      \inst{2}
      \and A. E. Chukwude
       \inst{2}
   }

   \institute{Department of Physics, Federal University of Technology Owerri, Imo state Nigeria\\
        \and
             Department of Physics and Astronomy, University of Nigeria, Nsukka, Nigeria\\
             \and
                 Department of Industrial Physics, Ebonyi State University, Abakaliki, Nigeria\\
{\it *romanusejike1971@gmail.com and romanus.ugwoke@futo.edu.ng}\\ 
\vs\no
   {\small Received~~20xx month day; accepted~~20xx~~month day}}

\abstract{We present the results of analyses of the ground level enhancements (GLEs) of cosmic ray (CR) events of 29 September 1989; 15 April 2001 and 20 January 2005. This involve examination of hourly raw CR counts of an array of neutron monitors (NMs) spread across different geographical latitudes and longitudes. Using awk script and computer codes implemented in R-software, the pressure corrected raw data plots of the NMs were grouped into low-, mid-, and, high-latitudes. The results show both similarities and differences in the structural patterns of the GLE signals. In an attempt to explain why the CR count during the decay phase of GLEs is always higher than the count before peak, we interpreted all counts prior to the peak as coming from direct solar neutrons and those in the decay phase including the peak as coming from secondary CR neutrons generated by the interactions of primary CRs with the atoms and molecules in the atmosphere. We identified NMs that detected these primary neutrons and found that they are close in longitudes. Previous authors seemingly identified these two species as impulsive and gradual events. Although there are a number of unexplained manifestations of GLE signals, some of the results  suggest that geomagnetic rigidity effectively determines the intensity of CRs at low- and mid-latitudes. Its impact is apparently insignificant in high-latitude regions. Nevertheless, the results presented should be validated before making any firm statements. Principally, the contributions of the ever-present and intractable CR diurnal anisotropies to GLE signals should be accounted for in future work. 
\keywords{Solar flare, Cosmic rays, Coronal mass ejection, ground level enhancement}
}

   \authorrunning{R.E. Ugwoke \& A.A. Ubachukwu \& J.O. Urama \& O. Okike \& J. Alhassan \& A.E. Chukwude} 
   \titlerunning{Analysis of GLE events of September 29, 1989; April 15, 2001 and January 20, 2005}  

   \maketitle

\nolinenumbers
%
\section{Introduction}           
\label{sect:intro}

Sharp or sudden increase in the intensity of cosmic rays (CRs) known as ground level enhancement (GLE) \citep[e.g.][]{Bk:2015,Fch:2010} arises from solar energetic particles (SEPs) with energies greater than 500MeV \citep{Yc:2015}.  
Reports from several authors have shown that the intensity of these relativistic particles is affected by a number of factors. A number of researchers \citep[e.g.][]{Belov:2009,teza:2016} indicate that GLEs frequently happened in association with Forbush decreases (FDs). This is expected as the two events mostly occur at the period of high solar activities. Though FDs and GLEs are interesting counterparts in the CR flux variations, a review of the available literature on the CR intensity variability shows that scientists investigating FDs have made significant advancements while those analyzing GLEs have recorded comparatively less progress. While both manual and automated detection of GLEs remains a challenge in the field, for example, timing and amplitude estimations of FD events have been successfully automated \citep[see][and the references therein]{ok:2020c,ok:2022,Alhassan:2022a}. While GLE event simultaneity has yet to be tested in detail, Forbush event global simultaneity have been the focus of several publications \citep{oh:08,oh:09,ok:2011,oh:2012,le:2015}. Location-dependent properties (e.g. rigidity, CR diurnal anisotropy, latitude, and longitude effects) of FDs have also been well investigated \citep[e.g.][]{teza:2016,ok:2019,ok:2019a} while there is a paucity of such detailed analyses for GLE phenomena. Using several NM stations, the current work seeks to examine the intensity variations of GLE events at different points at Earth.

Except for the primary neutrons, the geomagnetic field acts on the particles to limit the number that arrive on the surface of the earth in what is called geomagnetic cut-off rigidity ($R_{c}$)  \citep[e.g.][]{Ss:1991,Bchm:2013}. At any location on the surface of the earth, a SEP cannot arrive there if its rigidity does not exceed the $R_{c}$ value of that location. It is known that $R_{c}$ increases from the polar regions up to the equator where it is maximum. By implication, more of the SEPs arrive at the polar regions since the rigidity of the earth is low in these regions. On the other hand, few SEPs arrive at the equator because of the same reason. \citet{SS:01} studied five years interval of vertical cut-off rigidities for several CR stations from 1955 to 1995. Their results show that the $R_{c}$ values at any location is a changing variable \citep[see also][]{RS:19}. They noted that there were significant changes of $R_{c}$ in Latin America, South Africa, and near the coast of North America. The works of \citet{WI:18} and \citet{RS:19} suggest that computation of $R_{c}$ is directly linked to the asymptotic direction of approach of the SEPs. 
 
The trajectories through which CRs arrive from space can determine the intensity of CRs measured by NM. For each NM, this direction of arrival must be within a solid angle in the celestial sphere called asymptotic cone of acceptance \citep{Te:16}. The Tsyganenko96 model \citep[see][for example]{TS:89,PM:08} used in the determination of the asymptotic cone of acceptance of NMs show that there are contributions made by external magnetospheric sources and also  geomagnetic activity index (Kp index). The large differences in the variation of NM count rates at the time of GLEs are usually attributed to several other factors. Rather than blaming $R_{c}$, \citet{Bieber:2013}, for example, attributed the empirical differences in the magnitude of the peak intensities of GLEs at different locations of the Earth to atmospheric absorption, and thus, concluded that the observed significant differences in the NMs' count rates can be the result of anisotropy of CR particles. Therefore, to understand the reasons for the differences in the intensity of CRs recorded by NMs at close latitudes, one has to look beyond the geomagnetic cutoff rigidity.

A number of investigators \citep[see][for example]{ok:2011,Bt:2018,ok:2019,ok:2020b} observed that two or more NMs at close latitudes and longitudes could also have differing intensities because of the differences in their altitudes, NM sensitivities, temperature, detector type, spurious modulations, North-South anisotropy, snow, relative humidity, local wind speed and instrumental variations. Basically, air mass (and by extension, pressure) above a NM affects its count rates. The authors pointed out that besides pressure, NMs at high altitudes suffer disturbances from wind than others. This is most likely the reason why such NMs record lower intensity compared to their counterparts at the same latitude and longitude. The above number of competing factors makes the analysis of CR intensity variation (e.g. GLEs and Forbush decreases (FDs)) an interesting subject with numerous submissions \citep{si:54,lo:1969,Stroker:1994,Belov:2009,Usoskin:2011, be:2018b,Miroshnichenko:2018,ok:2019a,ok:2019b,ok:2019c} and many unexploited possibilities \citep[see][]{ok:2021d,ok:2021c,ok:2021,ok:2021b,Alhassan:2021,Alhassan:2022b, Alhassan:2022a, ok:2022}.

While some investigators \citep[e.g][]{teza:2016,teza:2016a,ok:2020c} employed a somewhat complicated method to demonstrate the implications of these spectacular phenomena (FDs and GLEs) on the latitudinal and longitudinal dependence of the amplitude of CR diurnal anisotropy, several others graphically presented the differences in the time-intensity profiles of GLEs at different locations of the Earth using NM CR count rates. Figure 1 of both \citet{Miroshnichenko:2000} and \citet{mcc:2008}, for example, respectively show the variability patterns of the GLE event of 29 September 1989 and that of 22 October 1989 using raw CR hourly data from two stations (McMurdo (MCMD) and Thule (THUL). Figure 1 of \citet{mcc:2008} shows that MCMD observes the largest CR flux increase for the event of 22 October 1989 whereas the GLE event of 29 September registers a relatively smaller increase at MCMD station. A pattern of variation, similar to the result of \citet{mcc:2008} is also presented in diagram 1 of \citet{Bieber:2013} for the GLE event of 20 January 2005. It is interesting to note here that THUL which records a very large increase for the event of 29 September only sees a very small increase for the event of 20 January 2005. These differences in the flux variations of the same events at different points at Earth are indications that GLEs might also be of different diversities \citep{Miroshnichenko:2018}; a reported characteristics of FDs \citep{be:08}. Rather than employing data from a few CR stations as is often the case in the literature, a detailed investigation of GLE manifestations requires analyses of GLEs events using data from an array of NM stations distributed around the Earth globe. 

.

In this article, we present a detailed study of the GLE events of 29 September 1989, 15 April 2001 and 20 January 2005. \citet{Asvstari:2017} report that GLEs are labeled following the event of 23 February 1956. This GLE happened first and was named GLE number 5 (apparently, there is no GLE number 1, 2, 3, or 4). Table 1 of \citet{Usoskin:2011} clearly shows that other GLEs are recorded/labeled sequentially GLE numbers 7, 8, 10, and so on. The GLE event of 29 September, 1989, 15 April 2001 and 20 January 2005 are respectively labeled GLE number 42, 60 and 69 \citep[see also][]{Mms:2012}. Table 1 contains the full names and the short names of the locations of the neutron monitors (NMs) used. Their latitudes and longitudes (in degrees) are in their respective plots. In this work, the short names are used as the name of the NMs.
\begin{table}
	\caption{Full names of the place where the neutron monitors are located and their short names.}
	\label{table:1}
	\centering
	\begin{tabular}{rll}
		\hline
		& NMFN & NMSN \\ 
		\hline
		1 & ALMAATAB & AATB \\ 
		2 & APATITY & APTY \\ 
		3 & BEIJING & BJNG \\ 
		4 & CALGARY & CALG \\ 
		5 & CAPE SHMIDT & CAPS \\ 
		6 & CLIMAX & CLMX \\ 
		7 & DEEP RIVER & DPRV \\ 
		8 & FORT SMITH & FSMT \\ 
		9 & GOOSEBAY & GSBY \\ 
		10 & HERMANUS & HRMS \\ 
		11 & INUVIK & INVK \\ 
		12 & IRKUTSK & IRKT \\ 
		13 & IRKUTSK\_2 & IRK2 \\ 
		14 & IRKUTSK\_3 & IRK3 \\ 
		15 & JUNGFRAUJOCH & JUN1 \\ 
		16 & JUNGFRAUJOCH\_2 & JUNG \\ 
		17 & KERGUELEN & KERG \\ 
		18 & KIEL & KIEL \\ 
		19 & KINGSTON & KGSN \\ 
		20 & LOMNICKY STIT & LMKS \\ 
		21 & MAGADAN & MGDN \\ 
		22 & MAWSON & MWSN \\ 
		23 & MCMURDO & MCMD \\ 
		24 & MEXICO CITY & MXCO \\ 
		25 & NAIN & NAIN \\ 
		26 & NEWARK & NWRK \\ 
		27 & NORILSK & NRLK \\ 
		28 & NOVOSIBIRSK & NVBK \\ 
		29 & OULU & OULU \\ 
		30 & POTCHEFSTROOM & PTFM \\ 
		31 & ROME & ROME \\ 
		32 & SANAE & SNAE \\ 
		33 & SOUTH POLE & SOPO \\ 
		34 & TBILISI & TBLS \\ 
		35 & TERRE ADELIE & TERA \\ 
		36 & THULE & THUL \\ 
		37 & TIXIE  BAY & TXBY \\ 
		38 & TSUMEB & TSMB \\ 
		39 & YAKUTSK & YKTK \\ 
		\hline
	\end{tabular}
\end{table}

\section{DATA SOURCE AND REDUCTION TOOLS}
\label{sect:data}
Hourly data was sourced from the World-Wide Neutron Monitor Network \url{http://cro.izmiran.ru/common/links.htm} hosted by the Pushkov Institute of Terrestrial Magnetism, Ionosphere, and Radio Wave Propagation, Russian Academy of Sciences (IZMIRAN). The data contains pressure corrected daily, hourly and sometimes minute CR count data.
Neutron monitor stations with a complete set of data covering a wide range of geomagnetic latitudes and longitudes were selected to enable one assess all the overall picture of the GLEs on earth for each event and possibly reflect peculiar responses of the monitors due to their locations.

Data from twenty-seven, twenty-two, and twenty-eight neutron monitors were analyzed for GLEs 42, 60 and 69 respectively. One hundred and twenty hours of observation were considered for GLEs 42 and 69 while seventy hours of observation were analyzed in the GLE 60 making it possible to assess the GCR counts prior to and after each GLE event.

Using awk programming, codes were written to transform the data from the website to data frames that can be read in R software package. Scripts implemented in R software environment were used first to plot the hourly CR counts for each station and for each event. Subsequent scripts implemented in the same R package plotted NM counts in groups based on their latitudes.

\section{RESULTS AND METHOD OF ANALYSIS }
\label{sect:results}
\subsection{RESULTS}
Figures 1 to 14 are the graphical results of this investigation. We employed an analytical method in the investigation. In doing so, we plotted the hourly CR count of each monitor with a complete set of data for each of the events under study. For the 29 September 1989 event (GLE 42), we grouped the NMs under six sections to include hourly CR ray counts at high-latitudes in the Northern hemisphere (latitude $60.50^{\circ}$ - $90.00^{\circ}$), upper mid-latitude in the Northern hemisphere (latitude $50.00^{\circ}$ - $60.00^{\circ}$), lower mid-latitude in the Northern hemisphere (latitude $31.00^{\circ}$ - $49.90^{\circ}$), low latitude ($-30.00^{\circ}$ - $30.00^{\circ}$), mid-latitude in the Southern hemisphere (latitude $-31.00^{\circ}$ - $59.90^{\circ}$) and high latitude in the Southern hemisphere (latitude $-60.50^{\circ}$ - $90.00^{\circ}$). Since there were more NMs in the mid-latitude Northern hemisphere we split them as mentioned above for better assessment.

The April, 15th 2001 event (GLE 60) was divided into four sub divisions: hourly CR counts at high latitudes in the Northern hemisphere (latitude $60.00^{\circ}$ - $90.00^{\circ}$), mid-latitude in the Northern hemisphere (latitude $31.00^{\circ}$ - $60.00^{\circ}$),  mid-latitude in the Southern hemisphere (latitude $-31.00^{\circ}$ - $59.90^{\circ}$) and high latitude Southern hemisphere (latitude $-60.50^{\circ}$ - $90.00^{\circ}$).

Similar to GLE 60, the January 20, 2005 event (GLE 69) was equally grouped into hourly CR counts at high latitudes in the Northern hemisphere (latitude $60.00^{\circ}$ - $90.00^{\circ}$), mid-latitude in the Northern hemisphere (latitude $31.00^{\circ}$ - $60.00^{\circ}$),  mid-latitude in the Southern hemisphere (latitude $-31.00^{\circ}$ - $59.90^{\circ}$) and high latitude in the Southern hemisphere (latitude $-60.50^{\circ}$ - $90.00^{\circ}$). These two last events did not have NM record of GLE in the low latitudes. This most probably is because they were weaker events when compared to GLE 42 and the $R_{c}$ is very high in this range of latitude.

The analysis after making group plots as pointed out above was to find out how the onset of the GLE started; specifically, one needed to know if there was an FD prior to the GLE in all the NMs or just some in any of the latitude groupings. The pattern of the CR counts was to be identified in terms of those having an instant rise to peak count before the decay and those who may have had two or more steps rise to the peak. 

The analysis was also intended to find out if there is a trend in the background GCR count with respect to the $R_{c}$ values at the location of the NMs. In addition, to find out what trend is there in the total count of the NMs with respect to not only the $R_{c}$ values but also their longitudes and altitudes.

For our results and analysis, we used the latitudes, longitudes and  vertical cut-off rigidities in table 1 of \citet{SS:01}: values for the 1990 were used in GLE 42  while the 1995 values were used in the GLEs 60 and 69 events. This is done with the hope that in the absence of the actual values at the time of these events, there was no significant variation in their values for each NM.


\section{Discussion}
\label{sect:discussion}
\subsubsection{Figure 1 Graph of the GLE 42 hourly CR count at high latitudes ($60.50^{\circ}$ - $90.00^{\circ}$) in the Northern hemisphere.}
Figure 1 is a combined plot of data from INVK, THUL, CAPS, OULU, APTY, MGDN and TXBY.  All the NMs in this group had the same onset phase. CAPS, THUL and MGDN are more closely related in having the same pattern of counting in all their phases. Their longitudes are also closer besides being in the same high latitude. They all had two steps to peak count which occurred in all of them at 12:00 UT.

It is our understanding that solar neutrons arrive undeflected by magnetic field both in the sun and on the earth. For this reason, hourly data will show them as the rising counts that may take two or more steps before the peak. It is possible that minute-by-minute data will show both their rising and decay (see \citet{DL:93}). The rest of what we have as the decay are secondary neutrons generated by the interaction of the primary protons with the atmosphere.

We also have APTY, OULU, and TXBY exhibiting the same pattern of counting in all their phases. They all had three-step rise to peak count and are found within the first quadrant of longitude. Having three steps imply that they had more of direct solar neutrons than their counterpart in this latitude range. INVK looks more like this group except that its count before the peak is more spaced than others. All the four NMs had their peaks at 13:00 UT.

MGDN with the highest $R_{c}$ recorded more counts than others. It also has the highest altitude (220m) and therefore should not have recorded more counts than those with zero altitudes (such as CAPS and OULU). Only the asymptotic cone of the NM can possibly account for this higher count it recorded. This is in line with \citep{Bedp:2004} and \citet{RS:19}, which show that at the polar region, $R_{c}$ is no longer the major determinant of the intensity of CRs as the asymptotic cone of the NMs.

INVK is the next to record highest count of CR after MGDN. Apart from THUL, it actually had the least R and is also closer to MGDN in longitude. CAPS has the next highest count after INVK.

The plot did not show that the GLE started at the recovery phase of an FD as pointed out in \citep[e.g.][]{Ukm:2011, Bk:2015}. In all the monitors too the recovery phase (in this case decay count) got back to the normal GCR the following day. 
\begin{figure}[htbp]
	\centerline{\includegraphics[width = 0.8\textwidth, angle=270]{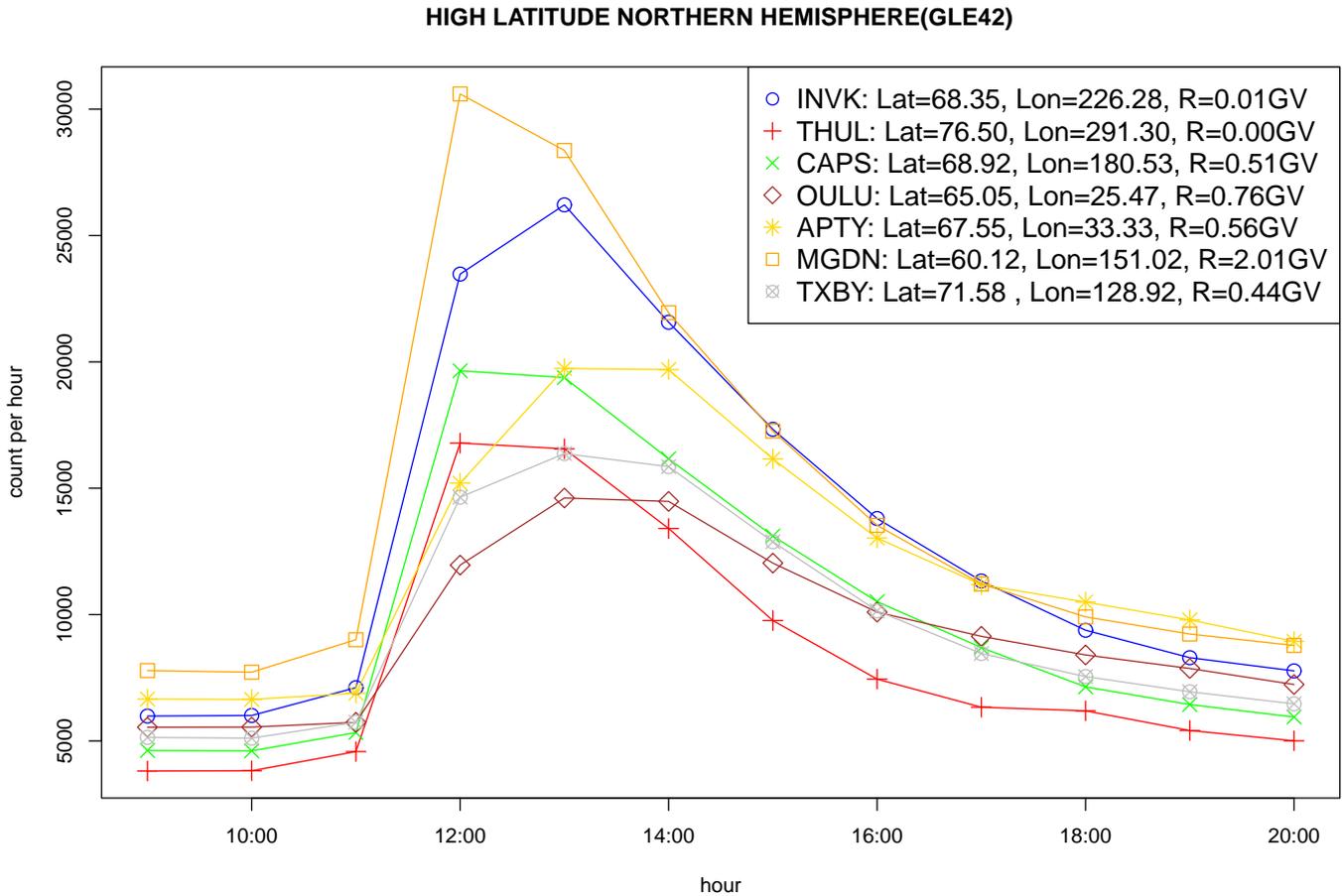}}
	\caption{Graph of the GLE 42 hourly CR count at high latitudes in the Northern hemisphere.}
	\label{}
\end{figure}

\subsubsection{Figure 2 Graph of the GLE 42 hourly CR count at upper mid-latitudes ($50.00^{\circ}$ - $60.00^{\circ}$) in the Northern hemisphere.}
In this group are IRK2, IRK3, IRKT, KIEL, CALG, NVBK and  GSBY with respective altitudes as 2000m, 3000m, 433m, 1008m, 1128, 163m, and 46m. The only common feature of the counting of CRs by these NMs is their similar onset that began with minute  increase in CR counts. This implies that they all recorded only fewer solar neutron counts at the onset.  
IRK2, KIEL and NVBK had their peaks by 13:00 UT while IRK3, IRKT and GSBY had theirs by 12:00 UT. There is no trend to determine why it is so. The pre-peak count recorded by CALG, NVBK, AND KIEL is an indication that they all had solar neutron counts just before counts due to solar proton generated secondary neutrons. 

Outside the initial small count of solar neutrons, IRKT, IRK2 and IRK3 did not record any more solar neutrons; instead, their peaks and the rest of their decay phase are solar proton-generated neutrons. This group had closer values of geomagnetic cut-off rigidities, latitudes and longitudes. The former group also shared closer values of cut-off rigidities than the others. The only one with an odd peak is GSBY whose cut-off rigidity is quite different.

In terms of total count up to the peak, CALG is the highest followed by NVBK and GSBY. If geomagnetic cut-off rigidity is the only reason for this, GSBY should have had more counts than CALG, however, CALG had counts about more than twice GSBY. The altitude is not equally the reason as the altitude of CALG (1128m) is much more than that of GSBY (46m). Higher altitude results in less count due to atmospheric (wind and snow) effects.

The very large increase observed by CALG for this event suggests a number of possibilities. It is most likely the longitude of CALG that led to its higher count. Though they are not in the high latitudes where it has been suggested in \citet{Bedp:2004} that $R_{c}$ is not effective in determining the intensity of NMs, it appears the same rule may apply at this range of latitude. This comparatively large increase may also be a pointer to sensitivity or high detection efficiency of the station \citep{ok:2020b}. The significantly larger increase may be specific to this event, and thus, an indication that the same GLE events may appear in different forms at different locations \citep{be:08,ok:2011}.  


\begin{figure}[htbp]
	\centerline{\includegraphics[width = 0.8\textwidth, angle=270]{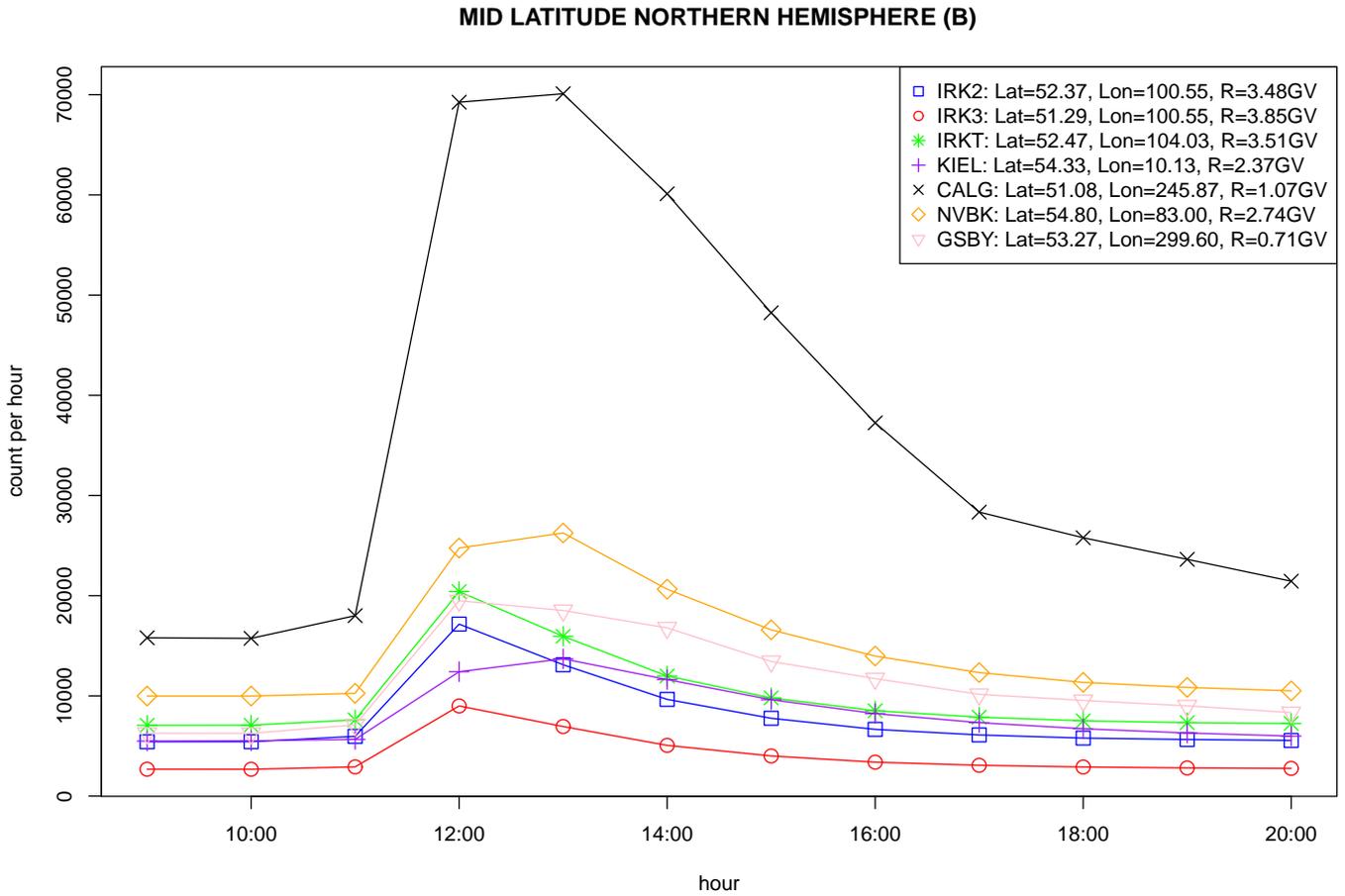}}
	\caption{Graph of the GLE 42 hourly CR count at upper mid-latitudes in the Northern hemisphere.}
	\label{}
\end{figure}  

\subsubsection{Figure 3 Graph of the GLE 42 hourly CR count at lower mid-latitudes ($31.00^{\circ}$ - $49.90^{\circ}$) in the Northern hemisphere.}
All the NMs in this group had a common pattern at the onset (a small rise in the CR count). This initial rise may be from solar neutrons.  JUN1 and JUNG are identical in all their phases. They are however more closely related to BJNG and TBLS in their pattern at the onset, peaks and decay phases. This is most likely due to their $R_{c}$ which are close in range to the other two. The four of them had a simultaneous peak at 12:00 UT which can be attributed to their close longitude apart from their close latitudes.

NWRK and DPRV are more related than others but the pre-peak count in DPRV is closer to the peak count than in NWRK where they are widely separated. This pre-peak count suggests the second arrival of groups of solar neutrons in the NMs. In comparison with those in Figure 2, their longitude is also close to that of CALG and GSBY  equally had such pre-peak counts. NWRK and DPRV had peaks at 13:00 UT probably due to closeness in their longitude. DPRV with the least $R_{c}$ had the highest count as expected. The peak counts and decay may arise from solar proton generated secondary neutrons.  The primary protons interact with the atoms and gasses in the atmosphere to generate the secondary neutrons \citep[e.g.][]{Mk:2002, Mms:2012,Bt:2018}.

BJNG equally had the least count as its $R_{c}$ is the highest. Thus unlike what we had in Figures 1 and 2, in this range of latitude, $R_{c}$ may play a significant role in the intensity recorded by NMs. It is not understood why NWRK with a low value of $R_{c}$ had a low count. Being closer to the equator like BJNG should have been enough reason for such count but its low $R_{c}$ calls to question if there was an error in the computation of $R_{c}$ by \citet{SS:01}.
\begin{figure}[htbp]
	\centerline{\includegraphics[width = 0.8\textwidth, angle=270]{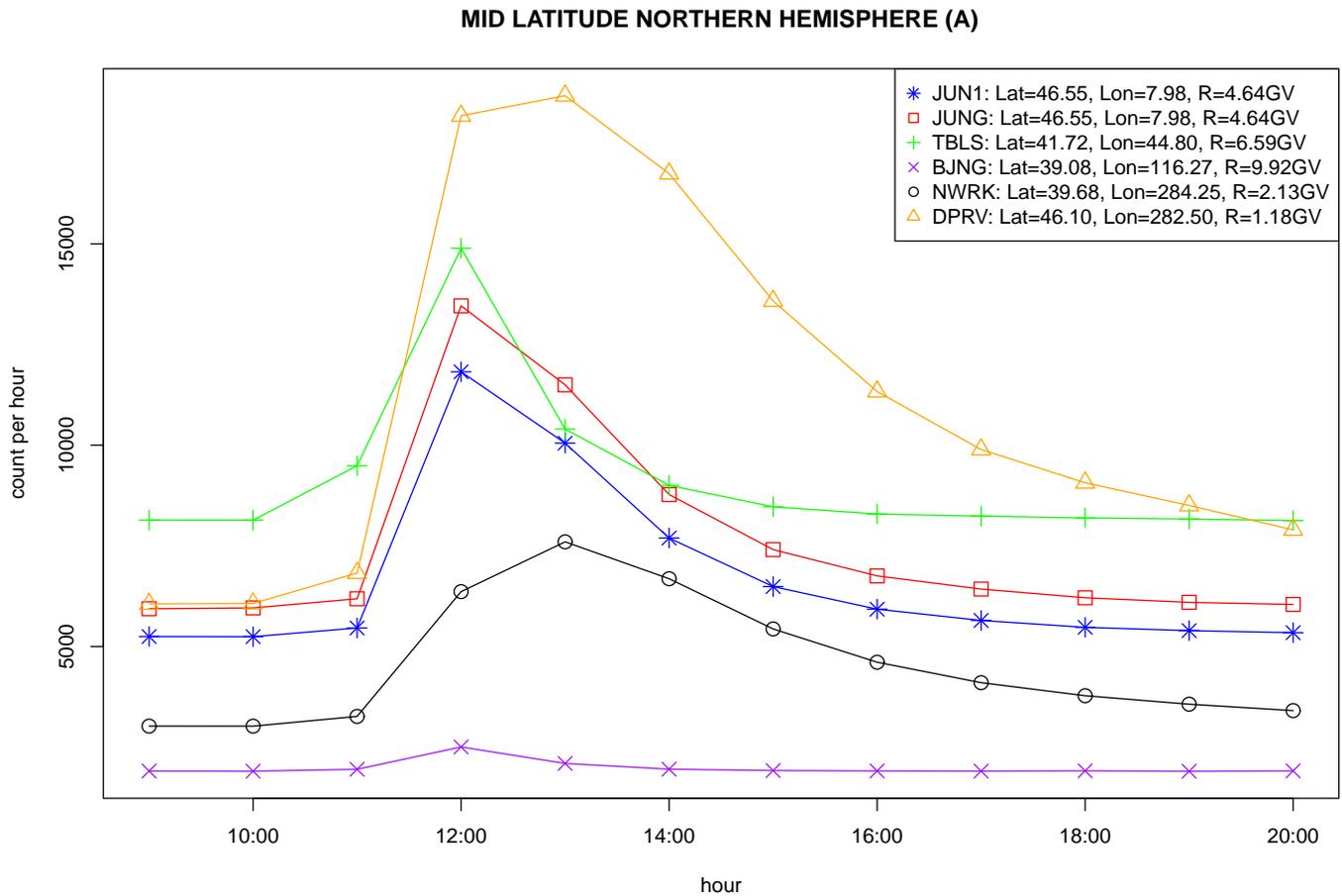}}
	\caption{Graph of the GLE 42 hourly CR count at lower mid-latitudes in the Northern hemisphere.}
	\label{}
\end{figure}

\subsubsection{Figure 4 Graph of the GLE 42 hourly CR count at mid-latitudes ($31.00^{\circ}$ - $60.00^{\circ}$) in the Southern hemisphere.}
KERG and HRMS both of which are in this group had similar onset that began with a small rise in CR count. HRMS peaked from this initial rise while KERG had a pre-peak count. This implies that their onset began with a few groups of a direct solar neutrons being recorded but only KERG had a second group of the same neutrons arriving at its location. 

On average, the background GCR count of KERG was much higher than that of HRMS. KERG having a greater number of total counts is consistent with its geomagnetic cut of  rigidity being lower than that of HRMS. Their altitudes could not have played any significant role in this difference in their CR count since their altitudes are relatively the same (33m for KERG and 26m for HRMS).

HRMS had its peak at 12:00 UT while KERG had its own at 13:00 UT. Their decay pattern was equally similar though their quasi-exponential decay began after the first drop in the CR count.
\begin{figure}[htbp]
	\centerline{\includegraphics[width = 0.8\textwidth, angle=270]{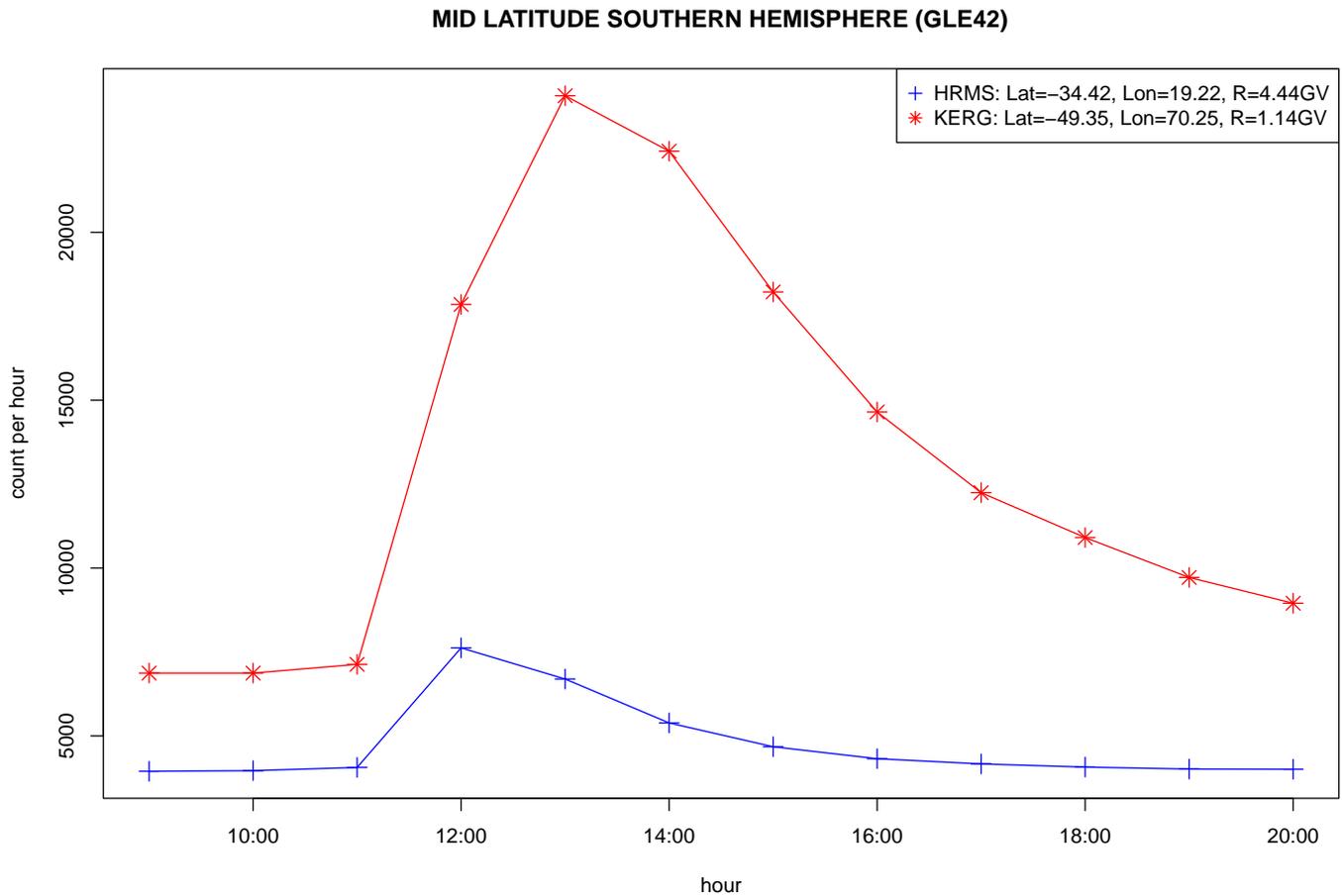}}
	\caption{Graph of the GLE 42 hourly CR count at mid-latitudes in the Southern hemisphere.}
	\label{}
\end{figure}

\subsubsection{Figure 5 Graph of the GLE 42 hourly CR count at high latitudes ($60.50^{\circ}$ - $90.00^{\circ}$) in the Southern hemisphere.}
In this group are SNAE, TERA and MCMD NMs. TERA had the least background GCR counts, followed by SNAE and then MCMD and this pattern correspond to their latitudinal positions (and subsequently their geomagnetic cut-off rigidities).
 
All of them had a similar patterns of CR counts in their onset, peak (14:00 UT for all of them) and decay phase. They all had three steps to their peak count which may not be visible in the group plot. Their onset began with a small increase in CR counts followed by the other steps. It also suggests that before the peak there were all of the arrival of a direct solar neutron in three batches; the first being much smaller than the others. Figure 5  also suggests that the greater the longitude, the more the groups of the solar neutrons recorded.

Their decay phase was quasi-exponential decay. All these related features could be because they share close values of latitude, longitude as well as altitudes ( SNAE=52m, TERA=32m, MCMD=48m).

Their differences lie in the background GCR count, at the onset, the total number of counts and the number of counts between each step. The background GCR count of TERA is on average 4,000 counts per hour and that of SNAE was about 6,000 counts per hour. MCMD which had the highest count has a background GCR count of about 8,000 counts per hour.
\begin{figure}[htbp]
	\centerline{\includegraphics[width = 0.8\textwidth, angle=270]{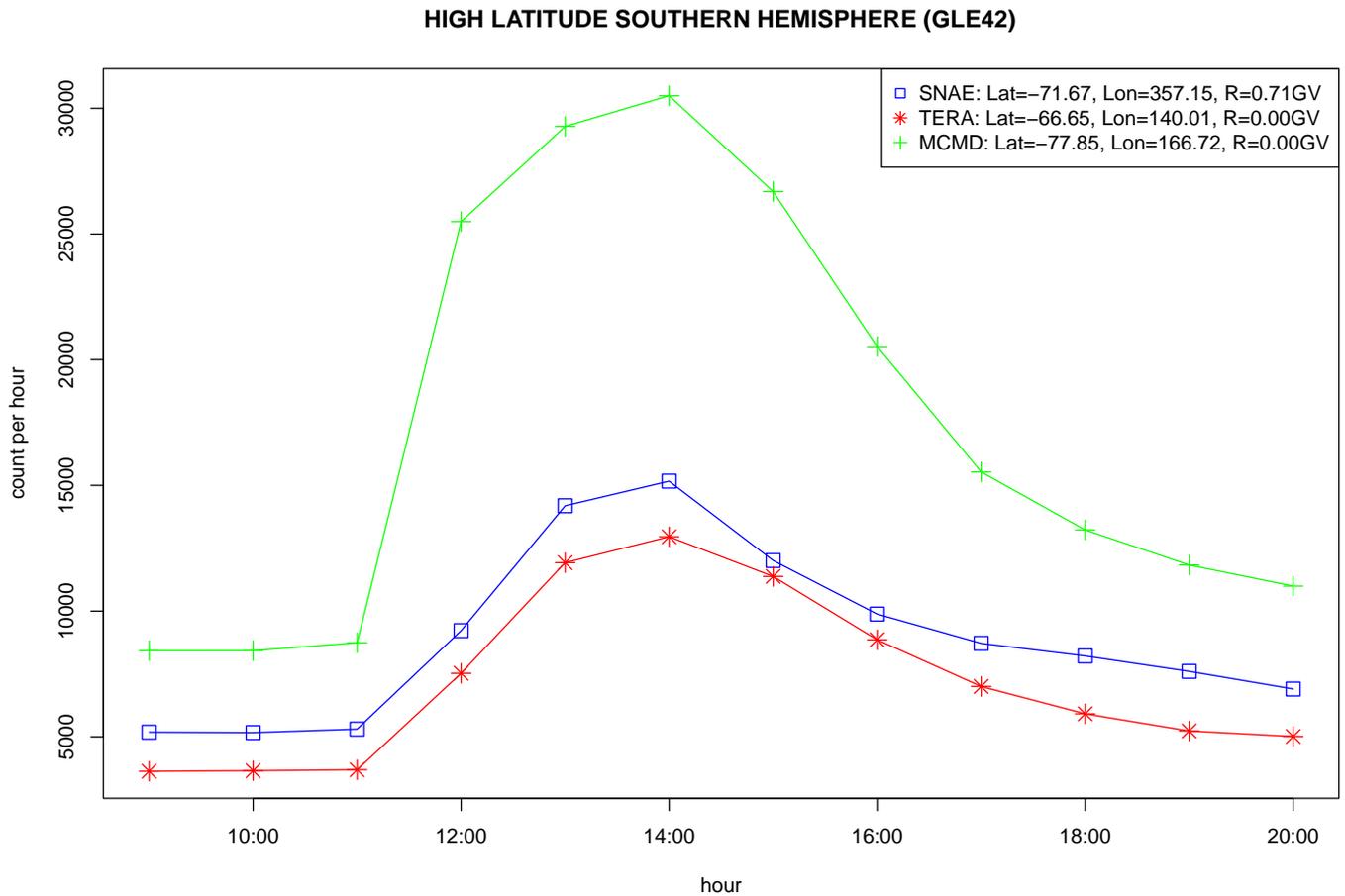}}
	\caption{Graph of the GLE 42 hourly CR count at high latitudes in the Southern hemisphere.}
	\label{}
\end{figure}

\subsubsection{Figure 6 Graph of the GLE 42 hourly CR count at low latitudes in the Northern hemisphere.}
In this group are PTFM and TSMB. The geomagnetic cut-off rigidity is usually high in this region (Latitude range $-30^{\circ}$ to $30^{\circ}$). Consequently, The NMs here usually record little count in CRs. The background GCR count by PTFM was as low as about 2,000 counts per hour on average and the total counts did not exceed 2,000 counts.

TSMB (Altitude=1240m) which was closer to the equator where the geomagnetic cut-off rigidity should be the highest had a greater background GCR count and total count than PTFM (1351m) . This can only be accounted for by their Altitudes as the atmospheric cut off is usually higher at greater Altitudes. 
 
While PTFM had its peak at 12:00, TSMB had its peak at 13:00 UT. This difference in the peak time may be due to the difference in their longitude. Both of them had initial onset at 10:00 UT that began with a small rise in CR count before rising to the peak count. There was not a clear FD prior to the GLE as shown in the plot. Their decay phase was a similar quasi-exponential drop in the CR count. There is nothing to suggest that they had a double peak.
\begin{figure}[htbp]
	\centerline{\includegraphics[width = 0.8\textwidth, angle=270]{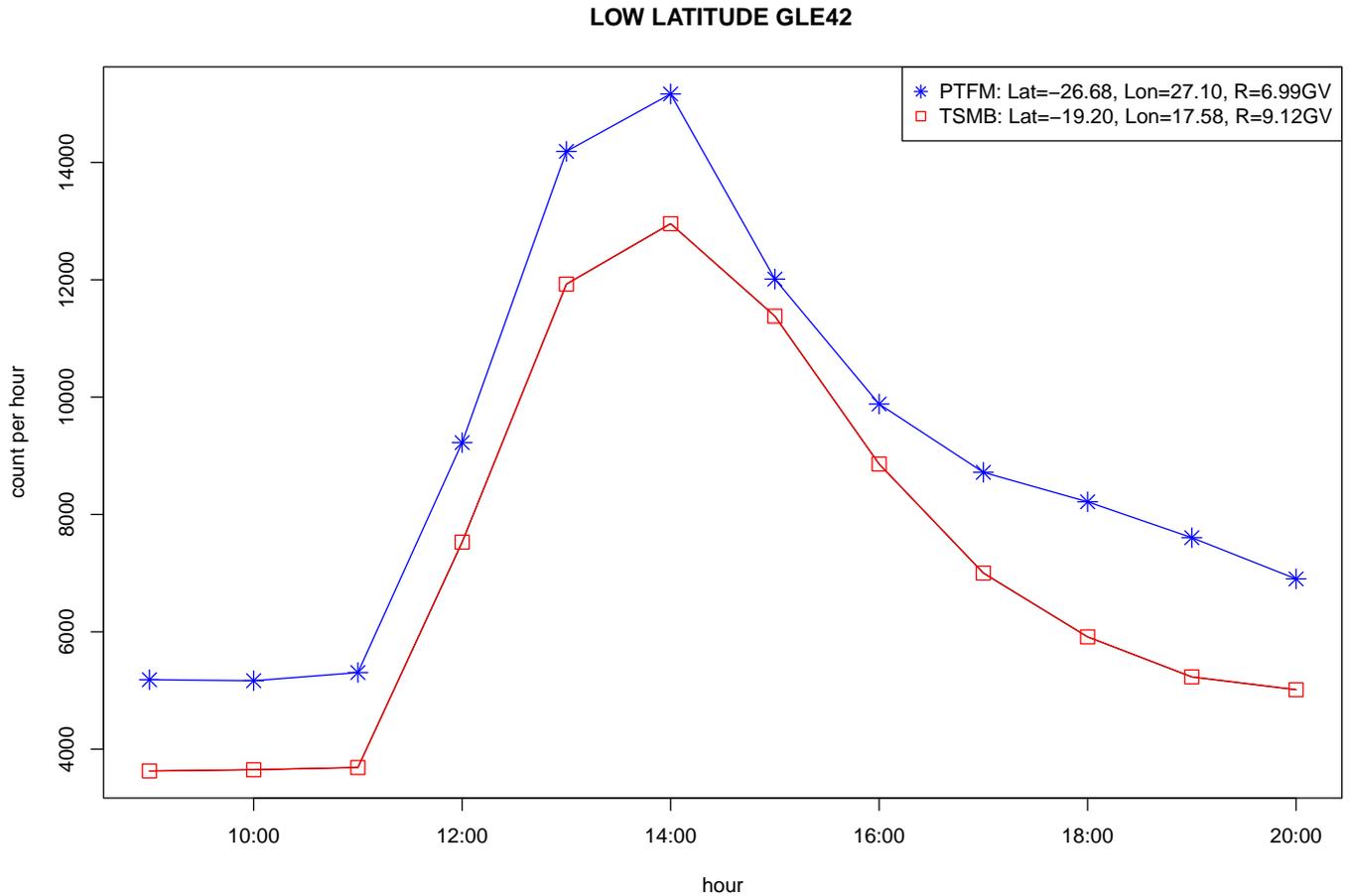}}
	\caption{Graph of the GLE 42 hourly CR count at low latitudes.}
	\label{}
\end{figure}

\subsubsection{Figure 7 Graph of the GLE 60 hourly CR count at high latitudes ($60.50^{\circ}$ - $90.00^{\circ}$) in the Northern hemisphere.}
All the NMs in this group had similar onset basically because they were in the same range of latitude ($60^{\circ}$ - $90^{\circ}$) and similar quasi-exponential decay. They however differed in their peaks.
Their onset suggests that GLE started at the recovery phase of a minor FD. They  may have recorded solar neutron counts at their onset. 

MGDN whose background GCR is supposed to be the least because of its high $R_{c}$ occupies the third position and also recorded the highest count. THUL is the least in terms of the background GCR yet its total CR count is the next highest. Whereas the $R_{c}$ value of THUL could be responsible for its high CR count, it is likely that the direction of viewing of MGDN or other factors led to its high CR count.


MGDN, OULU and THUL had an instant rise to the peak while NRLK, APTY, TXBY and CAPS had pre-peak counts. This implies that the first three in this group did not record any CR due to the direct solar neutron. They had their peaks earlier at 14:00 UT. The pre-peak counts of NRLk, APTY and CAPS are indications that they had recorded the arrival of the solar neutrons in their locations. This event has been listed  in \citet{Yc:2015} among those that had direct solar neutron associated with it. The direction of viewing of these NMs may have made it possible for them to view the solar neutrons. \citet{DL:93} have showed a similar situation in the May 24, 1990, CR event where only NMs in North America was said to be the only NMs that observed the direct solar neutrons. This gives credence to direction of viewing affecting the observation of NMs. 
 The NMs also had their peaks at 15:00 UT. In all of them, the R values did not follow the order of magnitude of their latitude and so their background GCR did not equally follow the order of the $R_{c}$ values. In this region too, MGDN with longitude $151.02^{\circ}$ has the greatest count and this is neither because of its altitude nor because of its $R_{c}$ value.
\begin{figure}[htbp]
	\centerline{\includegraphics[width = 0.8\textwidth, angle=270]{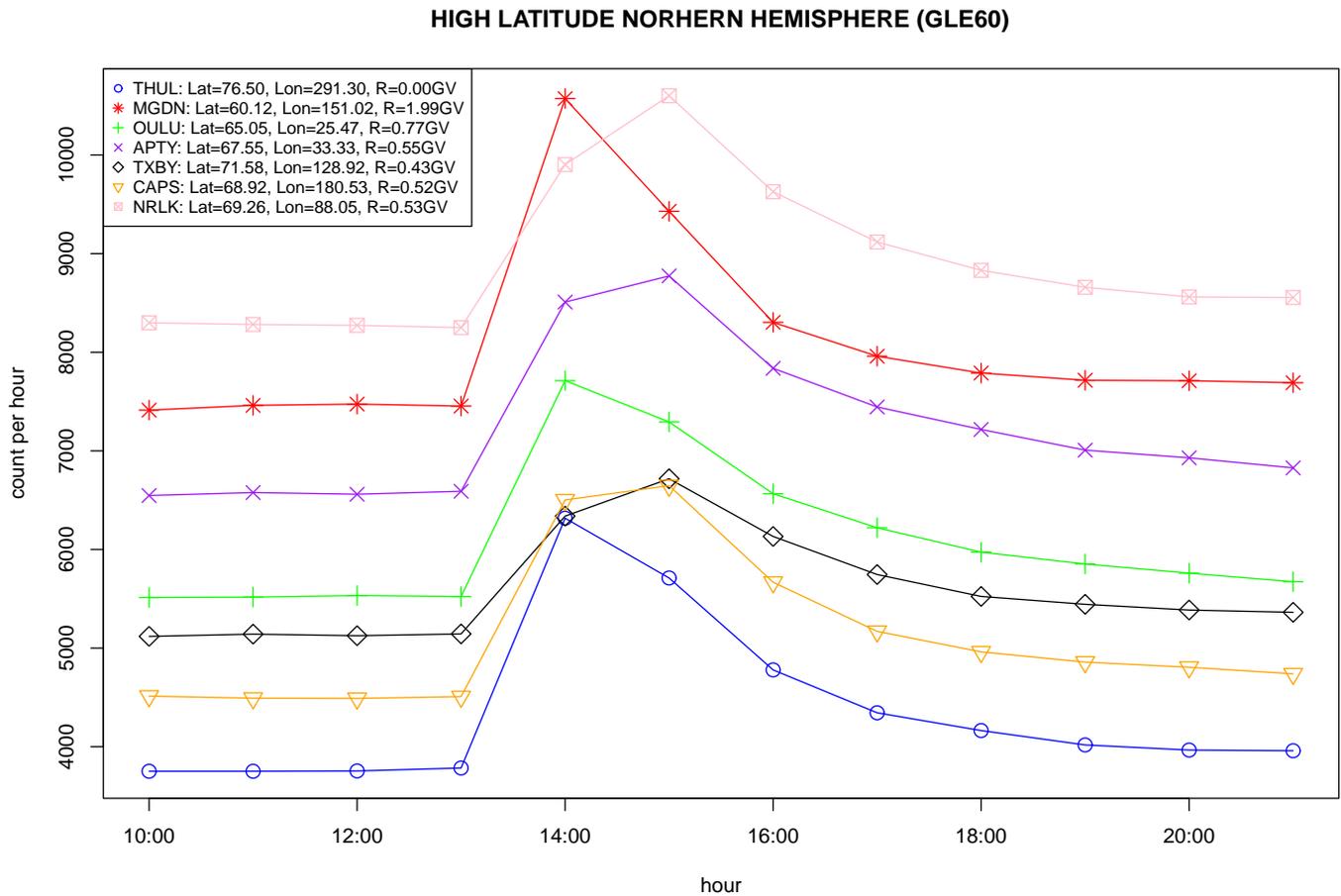}}
	\caption{Graph of the GLE 60 hourly CR count at high latitudes in the Northern hemisphere.}
	\label{}
\end{figure}

\subsubsection{Figure 8 Graph of the GLE 60 hourly CR count at mid-latitudes ($31.00^{\circ}$ - $60.00^{\circ}$) in the Northern hemisphere.}
NMs in this group include CALG, NVBK, IRKT, NAIN, JUNG, KIEL, IRK2, JUN1, and LMKS. Only NAIN and NVBK showed outstanding GLE and this has to do with the $R_{c}$ values being lower than that of others. Others had high values of R which by implication means that the earth's magnetic field restricts  CRs more at their locations. KIEL whose $R_{c}$ is comparable to that of NVBK did not record as much increase in CR count as NVBK possibly because of its direction of viewing (longitude). The first two with the highest CR counts (CALG and NAIN) had higher values of longitude. 

NAIN with $R_{c}$ = 0.45 is seen to have a lesser background GCR than NVBK with R= 2.69. This suggests that the number of CR particles available from the source also determines the CR count and not just the $R_{c}$. This does not mean that there may not be other reasons for such anisotropy in the GCRs.
\begin{figure}[htbp]
	\centerline{\includegraphics[width = 0.8\textwidth, angle=270]{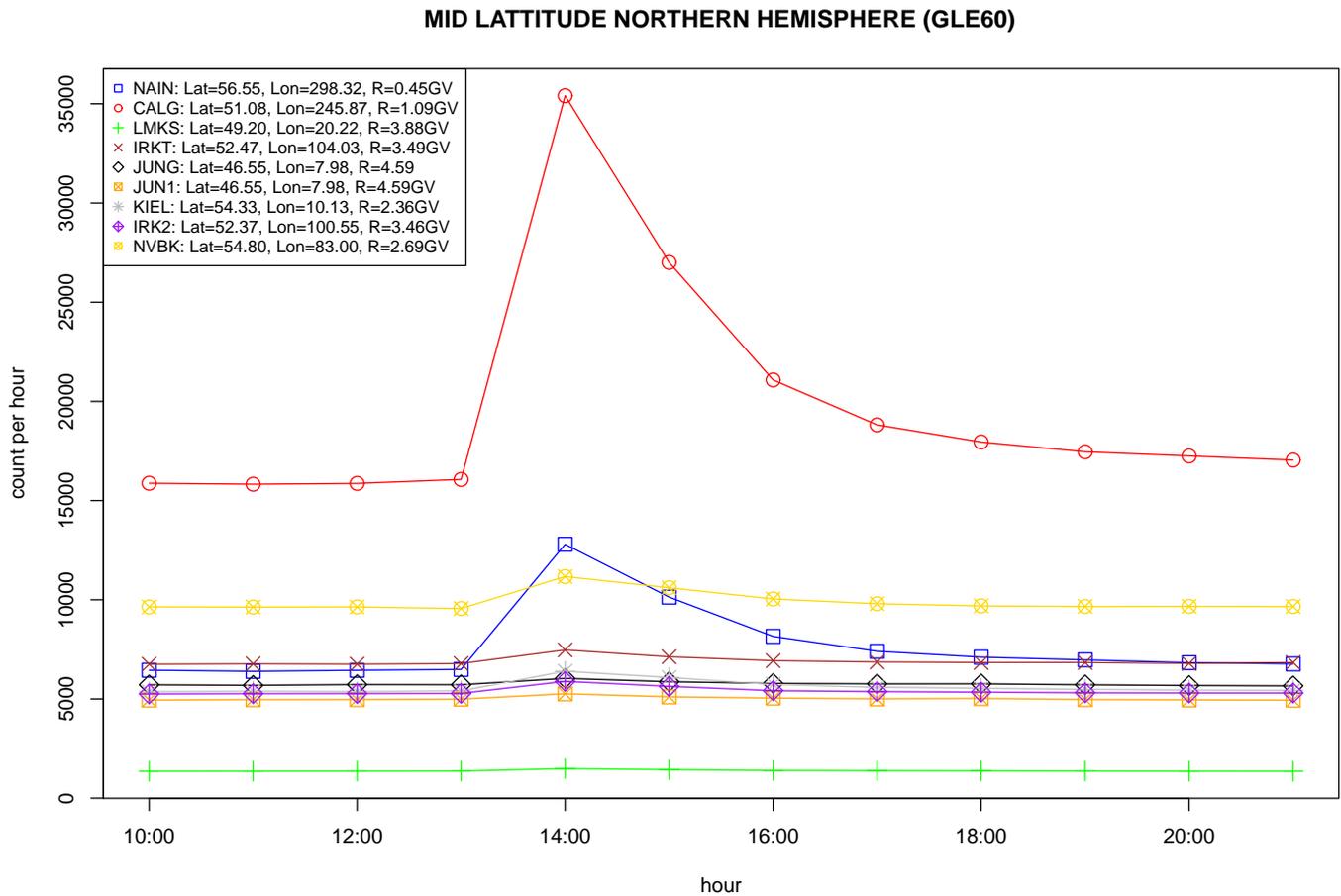}}
	\caption{Graph of the GLE 60 hourly CR count at mid-latitudes in the Northern hemisphere.}
	\label{}
\end{figure}

\subsubsection{Figure 9 Graph of the GLE 60 hourly CR count at mid-latitudes ($31.00^{\circ}$ - $60.00^{\circ}$) in the Southern hemisphere.}
Here are three NMs without any common feature in their CR counts. KGSN had an instant rise to the peak and recorded the highest count probably because of its direction of viewing or due to several other factors. KGSN may not record direct neutron from the sun. Its decay is just like that of KERG which is quasi-exponential. Its peak occurred at 14:00 UT.

KERG was almost like KGSN except that it had a pre-peak count and had its peak at 15:00 UT. It could have direct solar neutrons recorded as pre-peak count. KERG with a lower value of R$_{c}$ should have had a higher count but in this case, it did not. 
HRMS had the least count because of its high value of R$_{c}$. It also had its peak at 14:00 UT. Its initial phase was marked by series of rising and falling CR counts until it made a jump to peak count. This initial phase of rising and falling in CR counts might be due to the arrival of the solar neutrons in the NM. Its decay was not quasi-exponential like others. 
The latitude of HRMS appears to mark the boundary where the NMs stopped having a clear GLE. This is because all the NMs under study that are in this region (latitude $-30^{\circ}$ - $30^{\circ}$) recorded irregular CR counts that could not be justifiably said to be GLE.
\begin{figure}[htbp]
	\centerline{\includegraphics[width = 0.8\textwidth, angle=270]{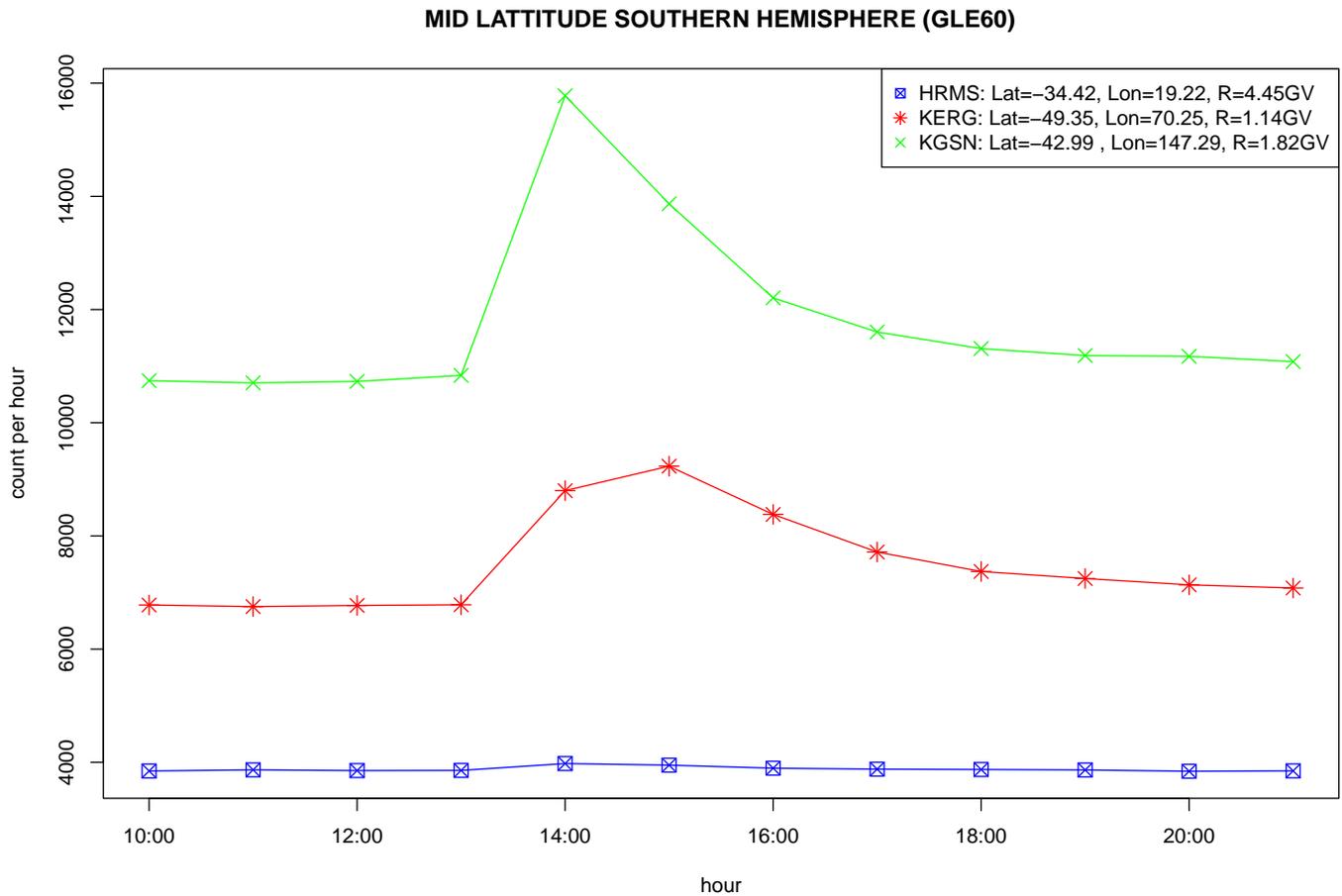}}
	\caption{Graph of the GLE 60 hourly CR count at mid-latitudes in the Southern hemisphere.}
	\label{}
\end{figure}

\subsubsection{Figure 10 Graph of the GLE 60 hourly CR count at high latitudes ($60.50^{\circ}$ - $90.00^{\circ}$) in the Southern hemisphere.}

SOPO, SNAE and MWSN are in this group. They all had similar onset and peak phases characterized by an instant rise in the peak count. The absence of any count before the peak is an indication that the lower nature of the magnitude of this event did not support much of direct solar neutrons. They also show that there was an FD whose recovery was part of the commencements of the enhancement. Their peaks simultaneously occurred at 14:00 UT. The recovery phase (or decay) never reverted to the initial background GCR rather it became higher. It appears that the final decay phase (or recovery phase) after GLE marks the beginning of the change in $R_{c}$ as seen in \citet{SS:01}. 

 SNAE had the highest count. This could not have been from its latitude or $R_{c}$ but rather from its direction of viewing, longitude other factors. Apparently, it appears that at the poles, high values of longitude (and or the asymptotic direction) played a dominant role in determining the NM with the highest count. This has been seen in GLE42 where MGDN had the highest count possibly for the same reason.
 \begin{figure}[htbp]
	\centerline{\includegraphics[width = 0.8\textwidth, angle=270]{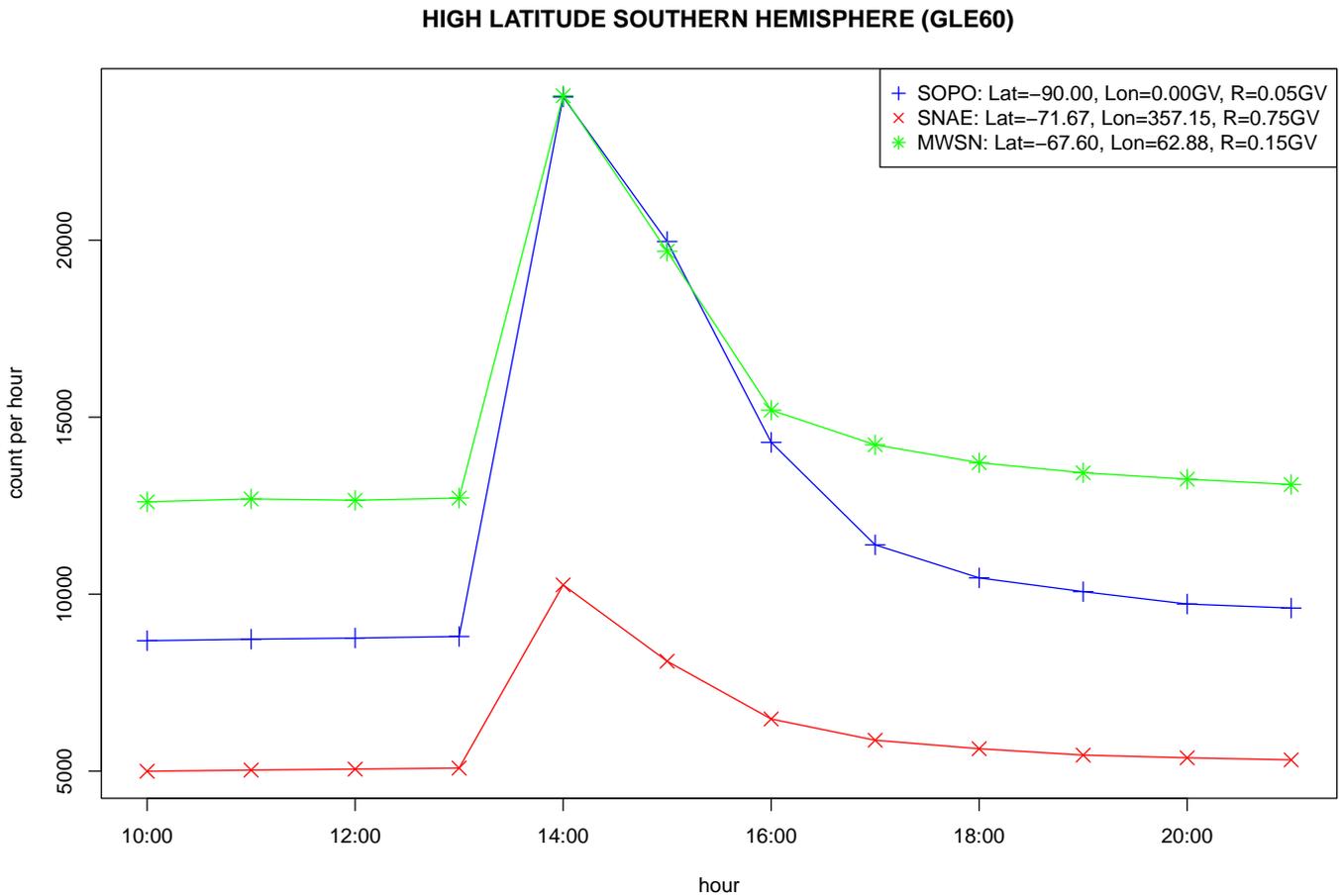}}
	\caption{Graph of the GLE 60 hourly CR count at high latitudes in the Southern hemisphere.}
	\label{}
\end{figure}

\subsubsection{Figure 11 Graph of the GLE 69 hourly CR count at high latitudes ($60.50^{\circ}$ - $90.00^{\circ}$  in the Northern hemisphere.}

In this group are three sub-groups: those that had an instant rise to peak count (such as MGDN, INVK,CAPS, and THUL), those that had a small rise in the CR count before the peak count (such as APTY, OULU, and YKTK) and NRLK which had a pre-peak count. Those that had instant rise to peak did not witness the arrival of direct solar neutrons in their locations. The second group had a record of a small number of group direct solar neutrons. NRLK also recorded this solar neutron arrival but the quantity in this case is much more in it than is in the second group. Though GLE 69 has not been specifically listed among GLEs associated with direct solar neutrons but studies made of most of the events that had direct solar neutrons associated with them shared some peculiar properties with GLE 69.

In \citet{Mms:2012}, it is seen that GLE69 had the short-lived component also called High energy impulsive ground level enhancement (HEIGLE). The HEIGLE lasted 20 minutes. The prompt or short-lived component of GLE 60 already known to be associated with direct solar neutron also lasted 14 minutes \citep{MMM:2008}. Also, according to \citet{CDF:1987}, the first increase of the June 3, 1982 event associated with direct solar neutron event did not last more than 20 minutes. Similarly, direct solar neutrons were said to be responsible for the prompt component (first increase) of the May 24, 1990 CR event \citep{DL:93}. Only NMs at North America were said to have observed the solar neutron part of the May 24, 1990 event. GLE 69 with all these similar properties should not be an exception to having direct solar neutrons. 

In NRLK there was an instant rise to a high value before the peak count. None of them manifested evidence of FD being at its recovery phase when the enhancement began. Their background GCRs appear to be relatively constant or flat. All the NMs had their peak count at 07:00 UT and each decay was quasi exponential. In all of them, the decay never returned to the earlier background GRC counts instead it is as if a new average background GCR count has been established.

Because of the simultaneity of their peak count, it is not easy to determine from the plot the order of the total count. APTY, however, had the highest count even though its $R_{c}$ value was not the least. THUL had the least count notwithstanding that its $R_{c}$ value shows that no CR is cut-off by the geomagnetic field where it is located. Its direction of viewing or other factors could be responsible.
\begin{figure}[htbp]
	\centerline{\includegraphics[width = 0.8\textwidth, angle=270]{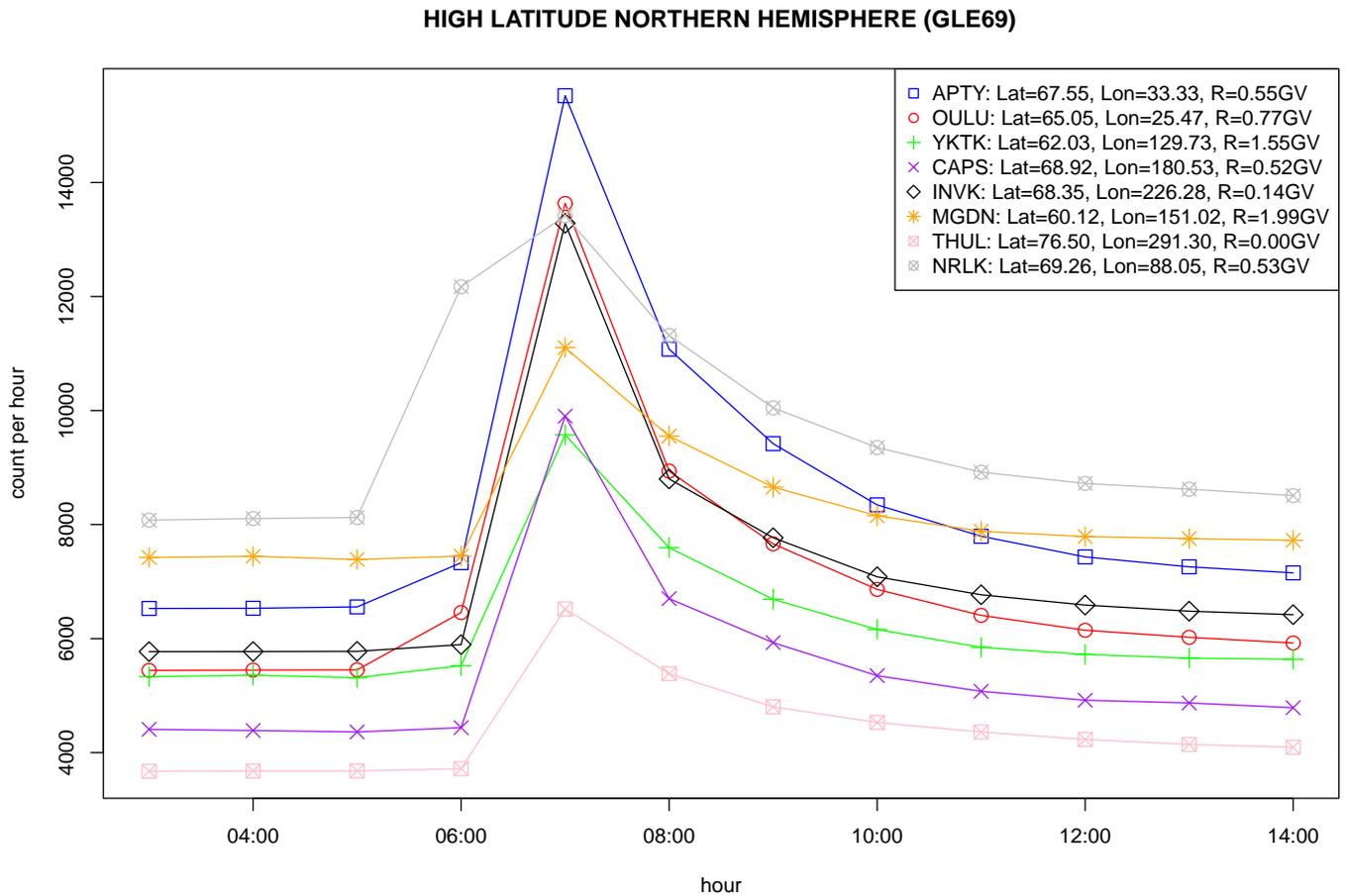}}
	\caption{Graph of the GLE 69 hourly CR count at high latitudes in the Northern hemisphere.}
	\label{}
\end{figure}

\subsubsection{Figure 12 Graph of the GLE 69 hourly CR count at mid-latitudes ($31.00^{\circ}$ - $60.00^{\circ}$) in the Northern hemisphere.}
All the NMs in this group showed an FD prior to the onset of the GLE. Except for NVBK, they all had another count before the peak which was a small rise in CR count at the point where the FD recovery phase terminated. This implies that all of them except NVBK had a count of a small number of solar neutrons. The order of the average background GCR was not based on the $R_{c}$ values of the NMs.

Only CLMX had peak count at 06:00 UT while the rest had theirs at 07:00 UT. Not all of them had quasi exponential decay like CALG. 

CALG had the highest count even more than NAIN with the least $R_{c}$ value. This is most likely due to its direction of viewing. NAIN with the least $R_{c}$ had the next total count. It is clear that its longitude is also close to that of CALG. However NAIN at longitude $-61.68^{\circ}$ is closer to longitude $0^{\circ}$ than CALG at longitude $245.87^{\circ}$ ($-114.13^{\circ}$). The next, CLMX and KIEL, were affected by their high $R_{c}$ values. Those with the highest $R_{c}$ value appears not to have had meaningful enhancements. These are JUN1, LMKS, IRK3 and IRK2. They were all characterized by a series of rising and falling CR counts until when they had a little increase followed by a decay that was not completed before another increase.

The plot for NMs at lower latitudes ($-30^{\circ}$ to $30^{\circ}$) were not included because they had irregular counts that could not be called GLE.
\begin{figure}[htbp]
	\centerline{\includegraphics[width = 0.8\textwidth, angle=270]{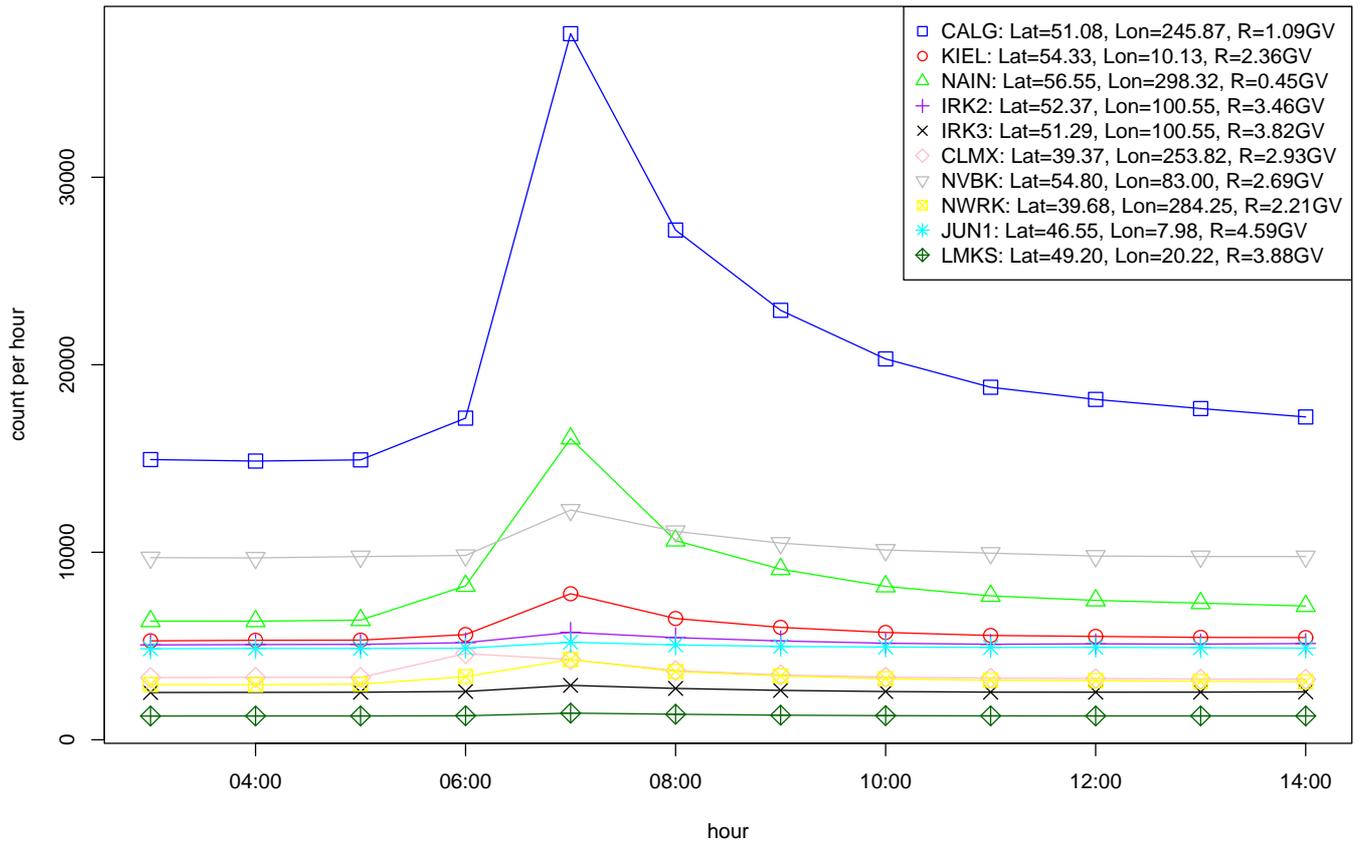}}
	\caption{Graph of the GLE 69 hourly CR count at mid-latitudes in the Northern hemisphere.}
	\label{}
\end{figure}

\subsubsection{Figure 13 Graph of the GLE 69 hourly CR count at mid-latitudes ($31.00^{\circ}$ - $60.00^{\circ}$) in the Southern hemisphere.}
In all the three NMs in this group, the onset, the peak and the decay pattern  were similar. They all appeared to have had an FD whose recovery phase marked the beginning of the GLE. After the FD recovery phase, there was first a small rise in CR count before a jump to the peak count. The small rise in the CR count is from the direct solar neutrons that arrived first because they suffered no deflection by the geomagnetic field. 

HRMS recorded a  very small rise in CR count not only because it was closer to the equator but its $R_{c}$ value was equally high. It, therefore, had a series of rising and falling CR counts before the GLE occurred. It did not fully decay to the background GCR before rising again.
In the three of them, the order of the background GCR count was according to their $R_{c}$ values where greater $R_{c}$ values correspond to lesser GCR counts. KGSN, however had more counts than KERG and this is not because of its altitudes but rather because of its longitude ($147.29^{\circ}$) that exposed it to favourable direction of viewing. Previous cases in this study (GLE42 and GLE60) where at a given range of latitude  NM with a high $R_{c}$ value recorded more counts also show that the atmospheric cut-off due to higher altitude did not cause the NM to have lesser count. Only in situations where two or more NMs with close values of latitude, longitude and $R_{c}$   did atmospheric cut-off (due to higher altitude) manifest.
Three of the NMs had their peaks at 07:00 UT.
\begin{figure}[htbp]
	\centerline{\includegraphics[width = 0.8\textwidth, angle=270]{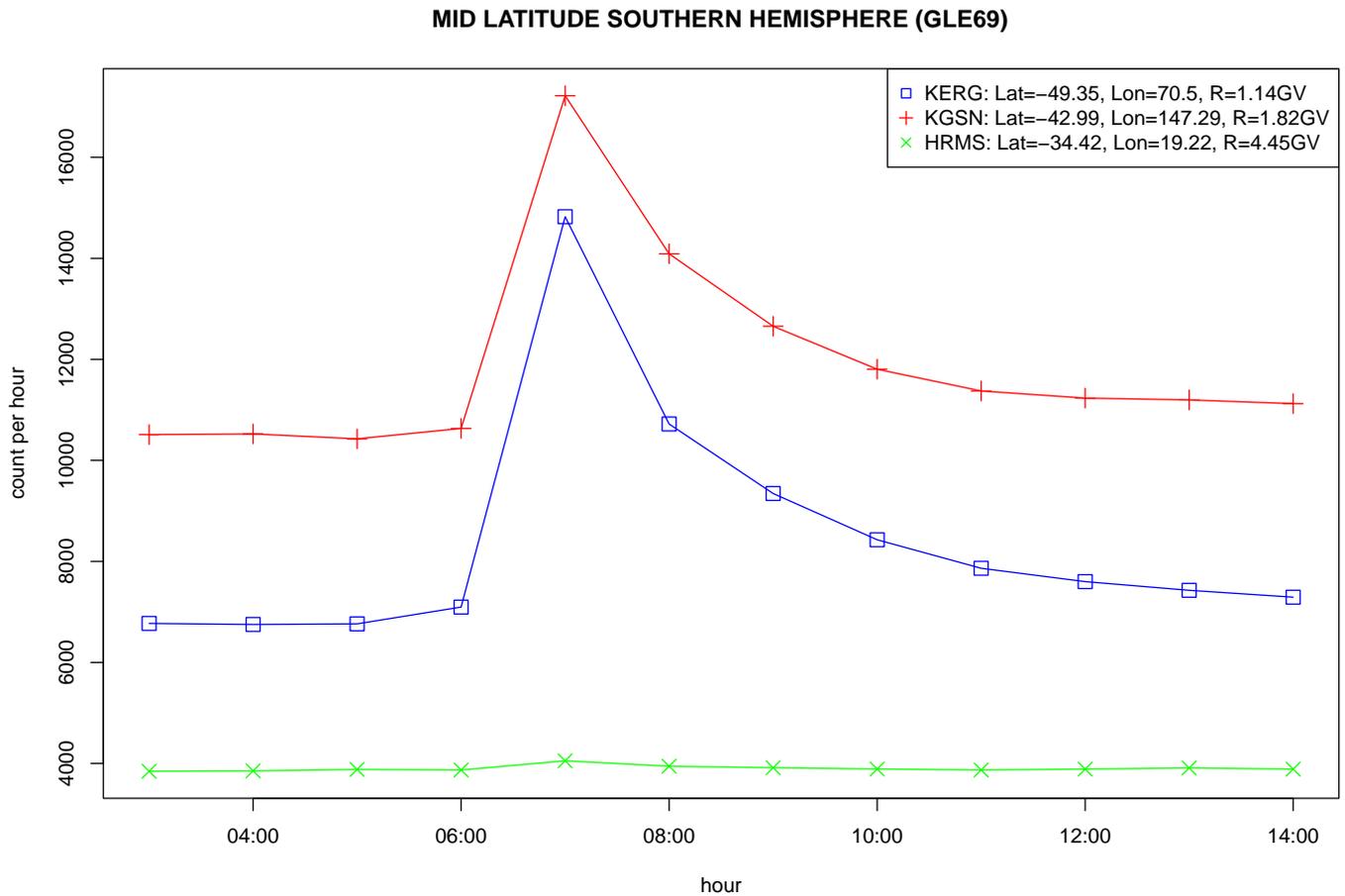}}
	\caption{Graph of the GLE 69 hourly CR count at mid-latitudes in the Southern hemisphere.}
	\label{}
\end{figure}

\subsubsection{Figure 14 Graph of the GLE 69 hourly CR count at high latitudes ($60.50^{\circ}$ - $90.00^{\circ}$)in the Southern hemisphere.}

In this group, SOPO and TERA showed no sign of prior FD before the GLE onset. They both had an instant rise to peak count. SNAE and MWSN showed there was an FD prior to the GLE. They equally had a small rise in CR count before a rapid rise to the peak count. SOPO and TERA also had a close resemblance in their decay just as SNAE and MWSN looked similar in their decay pattern.

\begin{figure}[htbp]
	\centerline{\includegraphics[width = 0.8\textwidth, angle=270]{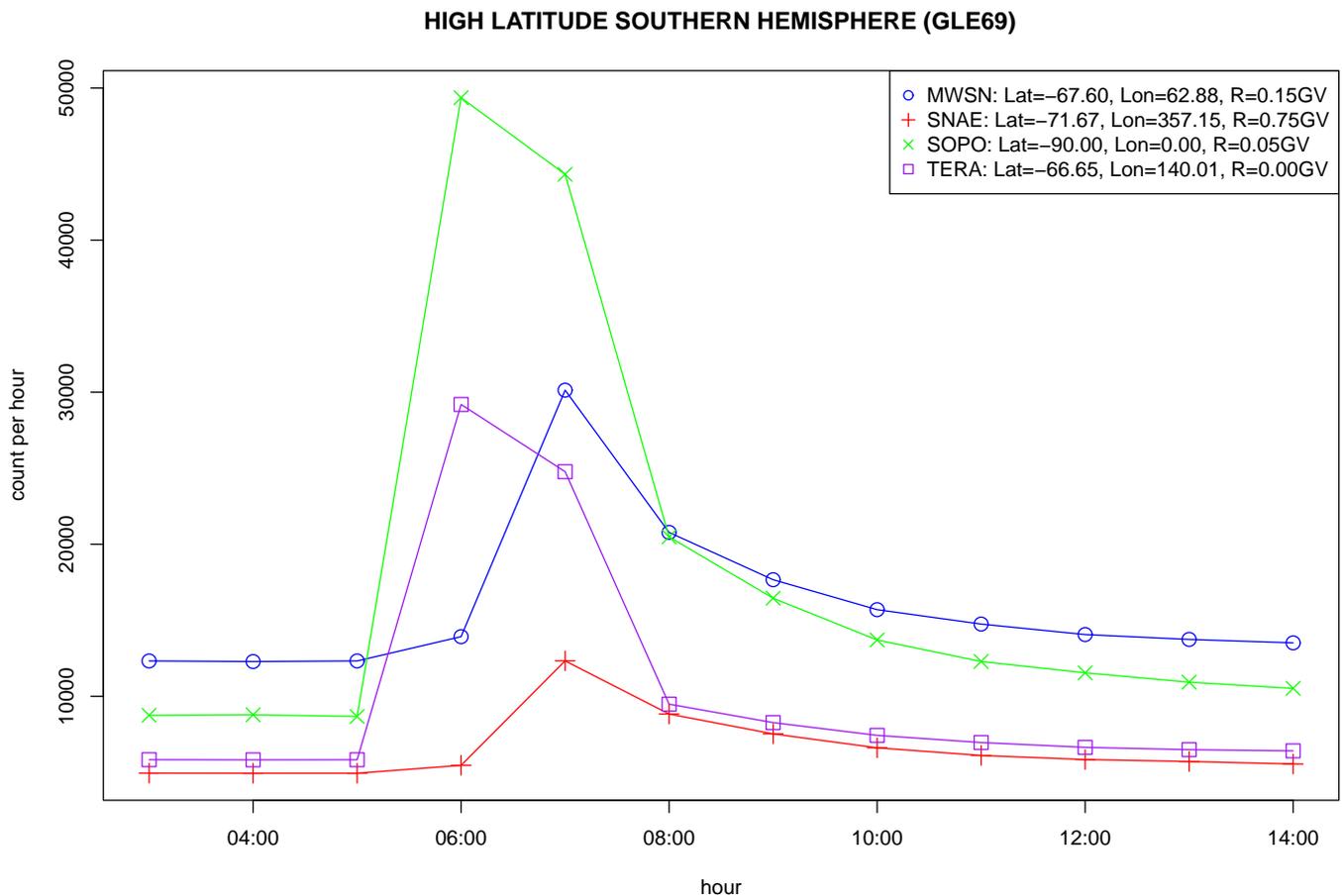}}
	\caption{Graph of the GLE 69 hourly CR count at High latitudes in the Southern hemisphere.}
	\label{}
\end{figure}

\subsubsection{Figure 15: Graph of Concurrent Forbush Decrease with GLE 69}
In Figure 15, we see a special case of conspicuous Forbush decrease at the time other monitors were recording enhancement in CR counts. The NMs involved are FSMT, MCMD and TXBY. \citet{Pmb:2010} analyzed a series of complex CR events that occurred between 17 January 2005 and 23 January 2005 using solar, interplanetary and, ground-based CR data. They reported that within the period, the Sunspot activities which generated a series of X and M flares with associated halos of CMEs gave rise to concurrent FDs and the GLE recorded on 20 January 2005. We see that in MCMD, when the FD was about to recover, there was another depression before it finally recovered and had enhancement from this recovery phase. This shows that the structure that formed the FD may not be one but are too close. The passage of the first was about to be completed before the second arrived. 

Since there is at least one station (CAPS) within this range of latitudes, longitudes and altitudes that did not observe such FD, it is likely that they could not see the FD because of their asymptotic cone of acceptance.
They were able to see enhancement after the FD. In all the three, both FD and GLE were conspicuous even though the latter was not as much as in the rest of the stations so far considered. They equally did not show any record of counting direct solar neutrons.
\begin{figure}[htbp]
	\centerline{\includegraphics[width = 0.8\textwidth, angle=270]{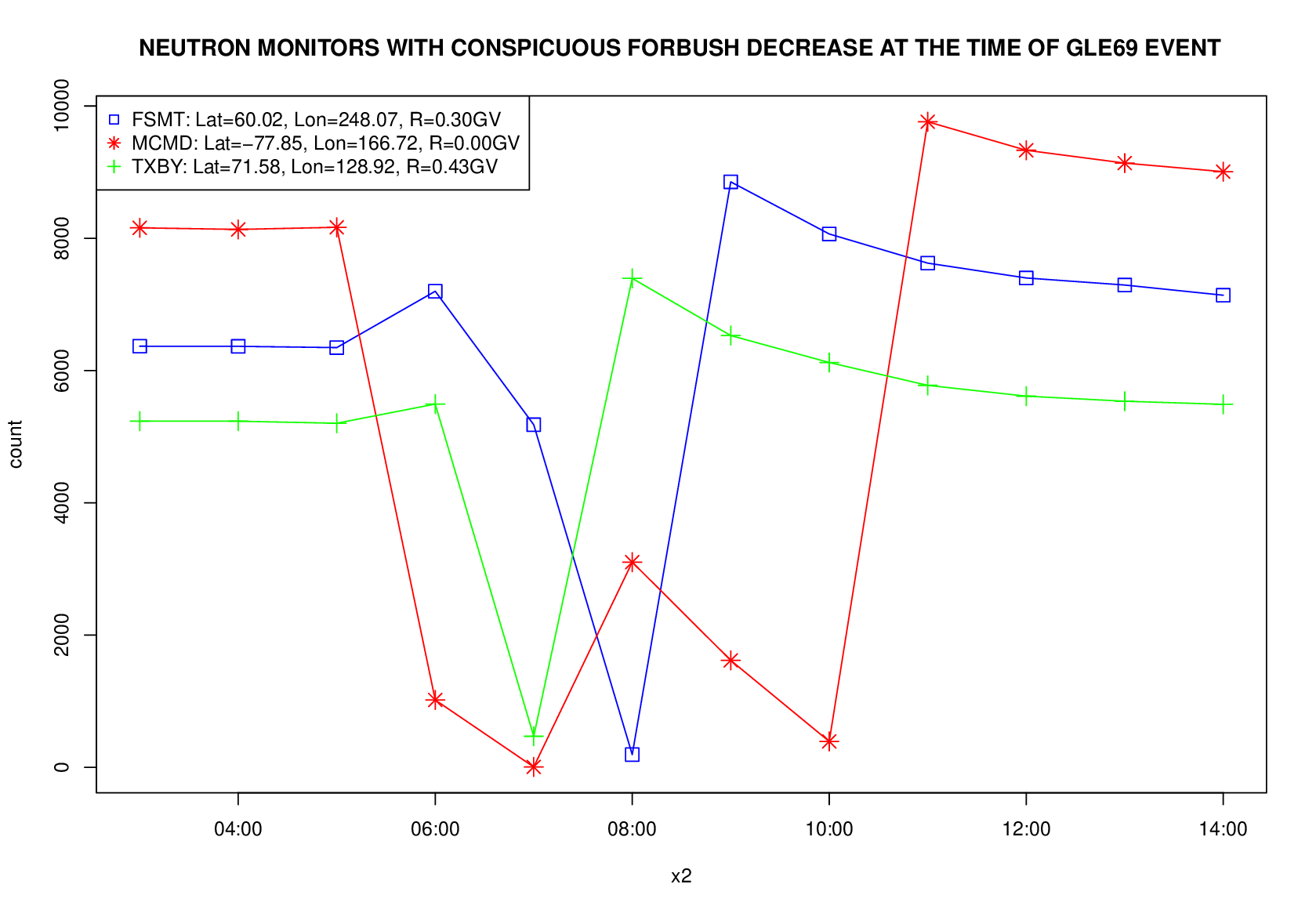}}
	\caption{Graph of Concurrent Forbush Decrease with GLE 69}
	\label{}
\end{figure}

\section{SUMMARY}
\label{sect:conclusion}
The detailed analyses of GLEs presented in the current work suggest that numerous factors affect CR measurements at the Earth. The general speculation that R$_{c}$, latitude and longitude do play a dominant role may be true as indicated by some of the presented results. Nonetheless, the differences in the manifestations and forms of GLEs at the stations examined underscore the need for the physics community to seek more inquiry into the causes of the phenomena. Indeed, some of the observed significant differences in the time-intensity profiles of GLEs at the different points at Earth make it hard to draw a firm conclusion regarding the properties of the three GLEs under study.

The difficulties in the analysis, interpretation and understanding of GLEs cast a strong question mark on the current traditional method of using raw CR data in the study of the phenomena. A large volume of publications \citep{ba:75,ca:96,cane:2003,Bieber:2013} submit that diurnal anisotropy contaminate CR measurements at Earth. \citet{be:08} and \citet{ok:2021c} found that the amplitude of CR diurnal anisotropy could even be larger than those of FDs or GLEs in some cases. It is desirable to untangle the effects of CR anisotropy before comparing the intensity profile of the events at different NM stations. This is because CR diurnal anisotropy could vary appreciably across the Earth and thus, influence their profiles significantly. One of the major flaws of the current work and other existing submissions on GLE characteristics is the use of raw or unprocessed \citep{ok:2020c} CR data. 

Before reaching a firm conclusion on the manifestations of GLEs and the factors that determine their characteristics at different latitudes and longitudes, an attempt should be made to validate the results presented here. We would pursue such a line of research in future work. The Fast Fourier Transformation method, recently developed by \citet{ok:2020c} and implemented in various applications \citep[e.g.][]{ok:2020,Alhassan:2022b}, would be used in future work to remove the effects of CR diurnal anisotropies on GLEs before comparing their structure at different NMs. Such investigation would not only be novel but also interesting. This is because some recent publications have empirically demonstrated that there are significant differences between time-intensity profiles/variations of raw and Fourier transformed CR data.

\section {Acknowledgments}
We are grateful to those hosting the website \url{http://cro.izmiran.ru/common/links.htm}. Their large data bank made our work easier. The efforts of the unknown reviewer is worth acknowledging. His/her input has a significant impact on the manuscript. We are indebted. We also wish to specially thank A. B. Belov for his personal communication. The list of neutron monitors he suggested was quite useful. Many thanks to the department of Physics and Astronomy, University of Nigeria Nsukka for providing the laboratory where we had to do our work. The Federal University of Technology Owerri Imo State Nigeria is acknowledged for helping with the Tartary Education Trust fund.


\bibliography{reference}

\begin{thebibliography}{57}
\providecommand\natexlab[1]{#1}
\providecommand\JournalTitle[1]{#1}

\bibitem[Alhassan {et~al.}(2021)]{Alhassan:2021}
Alhassan, J.~A., Okike, O., \& Chukwude, A.~E. 2021, RAA, 273

\bibitem[Alhassan {et~al.}(2022{\natexlab{a}})]{Alhassan:2022b}
Alhassan, J.~A., Okike, O., \& Chukwude, A.~E. 2022{\natexlab{a}}, Research in
  Astronomy and Astrophysics, 22,

\bibitem[Alhassan {et~al.}(2022{\natexlab{b}})]{Alhassan:2022a}
Alhassan, J.~A., Okike, O., \& Chukwude, A.~E. 2022{\natexlab{b}},
  J.Astrophys.Astr, 43

\bibitem[Asvestari {et~al.}(2017)]{Asvstari:2017}
Asvestari, E., Willamo, T., A., G., {et~al.} 2017, Advances in Space Research,
  60, 781

\bibitem[Badruddin \& Kumar(2015)]{Bk:2015}
Badruddin, \& Kumar, A. 2015, Solar Physics, 290, 4

\bibitem[Barouch \& Burlaga(1975)]{ba:75}
Barouch, E., \& Burlaga, L.~F. 1975, Journal of Geophysical Research, 80, 449

\bibitem[Belov(2008)]{be:08}
Belov, A.~V. 2008, Proceedings IAU Symposium, 257

\bibitem[Belov(2009)]{Belov:2009}
Belov, A.~V. 2009, International Astronomical Union, 439

\bibitem[Belov {et~al.}(2018)]{be:2018b}
Belov, A., Eronshenko, E., , {et~al.} 2018, Solar Physics, 293

\bibitem[Bieber {et~al.}(2004)]{Bedp:2004}
Bieber, J., Evenson, P., Droge, W., {et~al.} 2004, The Astrophysical Journal,
  601, 103

\bibitem[Bieber {et~al.}(2013)]{Bieber:2013}
Bieber, J.~W., Clem, J., Evenson, P., {et~al.} 2013, ApJ, 771

\bibitem[Blanco {et~al.}(2013)]{Bchm:2013}
Blanco, J., Catalan, E., Hidalgo, M., {et~al.} 2013, Solar Physics

\bibitem[Butikofer(2018)]{Bt:2018}
Butikofer, R. 2018, Springer, Cham, 1, 79

\bibitem[Cane \& Richardson(2003)]{cane:2003}
Cane, H.~V., \& Richardson, I.~G. 2003, JGR, 108, 1

\bibitem[Cane {et~al.}(1996)]{ca:96}
Cane, H.~V., Richardson, I.~G., \& {von Rosenvinge}, T.~T. 1996, Journal of
  Geophysical Research, 101, 21561

\bibitem[Chupp {et~al.}(1987)]{CDF:1987}
Chupp, E. L.and~Debrunner, H., Flückiger, E., {et~al.} 1987, The Astrophysical
  Journal, 318, 913

\bibitem[Derbrunner {et~al.}(1993)]{DL:93}
Derbrunner, H., Lockwood, J., \& Ryan, J. 1993, Apj., 409, 822

\bibitem[Firoz {et~al.}(2010)]{Fch:2010}
Firoz, K., Cho, K., Hwang, J., {et~al.} 2010, Journal of Geophysical Research,
  115, A09105

\bibitem[Lee {et~al.}(2015)]{le:2015}
Lee, S., Oh, S., Yu, Y., {et~al.} 2015, J, Geophys Res, 33, 33

\bibitem[Lockwood \& Webber(1969)]{lo:1969}
Lockwood, J.~A., \& Webber, W.~R. 1969, Journal of Geophysical Research, 74,
  5599

\bibitem[Martirosyan {et~al.}(2002)]{Mk:2002}
Martirosyan, H., Avakyan, K., Babayan, V., {et~al.} 2002, Report to COSPAR
  Congress, Houston, 2002, PSWI-CO.2-DO.1-E2.4-FO.1-PSRB2-0188-02

\bibitem[McCracken {et~al.}(2008)]{mcc:2008}
McCracken, K.~G., Moraal, H., \& Stroker, P.~H. 2008, JGR, 113, 1

\bibitem[McCracken {et~al.}(2012)]{Mms:2012}
McCracken, K., Moraal, H., \& Shea, M. 2012, The Astrophysical journal,
  761:101, 12

\bibitem[Miroshnichenko(2018)]{Miroshnichenko:2018}
Miroshnichenko, L.~I. 2018, J. Space Weather Space Clim, 8,

\bibitem[Miroshnichenko {et~al.}(2000)]{Miroshnichenko:2000}
Miroshnichenko, L.~I., De~Koning, C.~A., \& Perez-Enriquez, R. 2000, Space
  Science Reviews, 91, 615

\bibitem[Muraki {et~al.}(2008)]{MMM:2008}
Muraki, Y., Matsubara, Y., Masuda, S., {et~al.} 2008, Astropart.phys., 29, 229

\bibitem[Oh \& Yi(2012)]{oh:2012}
Oh, S.~Y., \& Yi, Y. 2012, Solar Phys, 280, 197

\bibitem[Oh \& Yi(2009)]{oh:09}
Oh, S., \& Yi, Y. 2009, Journal of Geophysical Research, 114

\bibitem[Oh {et~al.}(2008)]{oh:08}
Oh, S., Yi, Y., \& Kim, H.~Y. 2008, Journal of Geophysical Research, 113

\bibitem[Okike(2019{\natexlab{a}})]{ok:2019c}
Okike, O. 2019{\natexlab{a}}, The Astrophysical Journal, 882, 1

\bibitem[Okike(2019{\natexlab{b}})]{ok:2019a}
Okike, O. 2019{\natexlab{b}}, Journal of Geophysical Research: Space Physics,
  124, 1

\bibitem[Okike(2020{\natexlab{a}})]{ok:2020c}
Okike, O. 2020{\natexlab{a}}, JSTP, 211,

\bibitem[Okike(2020{\natexlab{b}})]{ok:2020}
Okike, O. 2020{\natexlab{b}}, MNRAS, 491, 3793

\bibitem[Okike(2021)]{ok:2021c}
Okike, O. 2021, ApJ, 60,

\bibitem[Okike \& Alhassan(2021)]{ok:2021d}
Okike, O., \& Alhassan, J.~A. 2021, Solar Phys, 296,

\bibitem[Okike \& Alhassan(2022)]{ok:2022}
Okike, O., \& Alhassan, J.~A. 2022, The European Physical Journal Plus, 137,

\bibitem[Okike {et~al.}(2021{\natexlab{a}})]{ok:2021b}
Okike, O., Alhassan, J.~A., Iyida, E.~U., \& Chukwude, A.~E.
  2021{\natexlab{a}}, MNRAS, 503, 5675

\bibitem[Okike \& Collier(2011)]{ok:2011}
Okike, O., \& Collier, A.~B. 2011, Journal of Atmospheric and Solar-Terrestrial
  Physics, 73, 796

\bibitem[Okike \& Nwuzor(2020)]{ok:2020b}
Okike, O., \& Nwuzor, O.~C. 2020, MNRAS, 493, 1948

\bibitem[Okike {et~al.}(2021{\natexlab{b}})]{ok:2021}
Okike, O., Nwuzor, O.~C., Odo, F.~C., {et~al.} 2021{\natexlab{b}}, MNRAS, 502,
  300–312

\bibitem[Okike \& Umahi(2019{\natexlab{a}})]{ok:2019b}
Okike, O., \& Umahi, A.~E. 2019{\natexlab{a}}, Journal of Atmospheric and
  Solar-Terrestrial Physics, 189, 35

\bibitem[Okike \& Umahi(2019{\natexlab{b}})]{ok:2019}
Okike, O., \& Umahi, A.~E. 2019{\natexlab{b}}, Solar Physics, 294

\bibitem[Papaioannou {et~al.}(2010)]{Pmb:2010}
Papaioannou, O., Malandraki, O., Belove, A., {et~al.} 2010, Solar Phys, 266,
  181

\bibitem[Plainaki {et~al.}(2008)]{PM:08}
Plainaki, C., Mavromichalaki, H., Belov, A., Eroshenko, E., \& Yanke, V. 2008,
  Advances in Space Research

\bibitem[Rosen \& Seunarine(2019)]{RS:19}
Rosen, L., \& Seunarine, S. 2019, Proceedings of Science, 1148

\bibitem[Simpson(1954)]{si:54}
Simpson, J.~A. 1954, Physical Review, 94, 426

\bibitem[Smart \& Shea(2001)]{SS:01}
Smart, D., \& Shea, M. 2001, Proceedings of ICRC 2001: 4063 c Copernicus
  Gesellschaft 2001, 4063

\bibitem[Smart {et~al.}(1991)]{Ss:1991}
Smart, D., Shea, M., Wilson, M., {et~al.} 1991, Proc. International Cosmic Ray
  Conference Dublin Ireland. SH 3.1-3, 3, 1

\bibitem[Stroker(1994)]{Stroker:1994}
Stroker, P.~H. 1994, Space Science Reviews, 73, 327

\bibitem[Tezari \& Mavromichalaki(2016)]{teza:2016a}
Tezari, A., \& Mavromichalaki, H. 2016, New Astron, 46, 78

\bibitem[Tezari {et~al.}(2016{\natexlab{a}})]{teza:2016}
Tezari, A., Mavromichalaki, H., Katsinis, D., {et~al.} 2016{\natexlab{a}},
  Annales Geophysicae, 34, 1053

\bibitem[Tezari {et~al.}(2016{\natexlab{b}})]{Te:16}
Tezari, A., Mavromichalaki, H., Katsinis, D, {et~al.} 2016{\natexlab{b}}, Ann.
  Geophys., 34, 1053..1068

\bibitem[Tsyganenko(1989)]{TS:89}
Tsyganenko, N. 1989, Planet Space Sci., 37 N0 1, 5..20

\bibitem[Usoskin {et~al.}(2011{\natexlab{a}})]{Usoskin:2011}
Usoskin, I.~G., Kovaltsov, G.~A., Mironova, I.~A., Tylka, A.~J., \& Dietrich,
  W.~F. 2011{\natexlab{a}}, Atmos. Chem. Phys, 11, 1979

\bibitem[Usoskin {et~al.}(2011{\natexlab{b}})]{Ukm:2011}
Usoskin, I., Kovaltsov, G., Mironova, I., {et~al.} 2011{\natexlab{b}}, Atmos.
  Chem. Phys., 11, 1979

\bibitem[Wozniak {et~al.}(2018)]{WI:18}
Wozniak, W., Iskra, K., Modzelewska, R., {et~al.} 2018, Researchgate Conference
  Paper · January 2018

\bibitem[Yu {et~al.}(2015)]{Yc:2015}
Yu, X., Lu, H., Chen, G., {et~al.} 2015, New Astronomy, 39, 25

\end{thebibliography}

\end{document}